\documentclass[10pt,twocolumn,journal]{IEEEtran}

\usepackage{graphicx}
\usepackage[english]{babel}
\usepackage{amssymb}
\usepackage{amsbsy}
\usepackage{multirow}
\usepackage{slashbox}
\usepackage{colortbl}
\usepackage{threeparttable}
\usepackage{algorithm}
\usepackage{algorithmic}

\title{Turbo NOC: a framework for the design of Network On Chip based turbo decoder architectures}

\author{Maurizio Martina, \emph{Member IEEE}, Guido Masera, \emph{Senior Member IEEE} 
\thanks{The authors are with 
Dipartimento di Elettronica - Politecnico di Torino - Italy. This work is partially supported by the WIMAGIC project and 
by the Newcom++ network of excellence, funded by the European Community.}}

\begin{document}

\maketitle

\begin{abstract}
This work proposes a general framework for the design and simulation of 
network on chip based turbo decoder architectures. 
Several parameters in the design space are investigated, namely the 
network topology, the parallelism degree, the rate 
at which messages are sent by processing nodes 
over the network and the routing strategy. 
The main results of this analysis are: i) the most suited topologies 
to achieve high throughput with a limited complexity overhead 
are generalized de-Bruijn and generalized Kautz topologies; 
ii) depending on the throughput requirements different parallelism degrees, 
message injection rates and routing algorithms can be used to minimize the network area overhead.
\end{abstract}


\section{Introduction}
\label{sec:intro}

In the last years wireless communication systems coped with the problem of delivering reliable information while 
granting high throughput. This problem has often been faced resorting to channel codes able to 
correct errors even at low signal to noise ratios. 
As pointed out in Table I in \cite{wehn_TVLSI08}, several standards for wireless communications adopt binary 
or double binary turbo codes \cite{berrou_ICC93, berrou_ITW01} and exploit their excellent error correction capability. 
However, due to the high computational complexity required to decode turbo codes, optimized architectures  
(e.g. \cite{dobkin_TVLSI05}, \cite{martina_TCASII08}) have been usually employed. 
Moreover, several works addressed the parallelization of turbo decoder architectures to achieve higher throughput. 
In particular, many works concentrate on 
avoiding, or reducing, the collision phenomenon that arises with parallel architectures 
(e.g. \cite{giulietti_EL02, lee_EL02, wehn_ISCAS02, tarable_CL04}).

Although throughput and area have been the dominant metrics driving the optimization of turbo decoders,
recently, the need for flexible systems able to support different operative 
modes, or even different standards, has changed the perspective. In particular, the so called software defined radio 
(SDR) paradigm 
made flexibility a fundamental property \cite{polydoros_PIMRC08} of future receivers, which will be 
requested to support a wide range of heterogeneous standards.
Some recent works  
(e.g.  \cite{wehn_TVLSI08}, \cite{bougard_ICT08}, \cite{baghdadi_TVLSI09}) deal with 
the implementation of Application-Specific Instruction-set 
Processor (ASIP) architectures for turbo decoders. In order to obtain architectures that achieve both high throughput 
and flexibility multi-ASIP is an effective solution. Thus, together with flexible and 
high throughput processing elements, a multi-ASIP architecture must feature also a flexible and high throughput 
interconnection backbone. To that purpose, 
the Network-On-Chip (NOC) approach has been 
proposed to interconnect processing elements in turbo decoder architectures 
designed to support multiple standards 
\cite{wehn_icecs02}, \cite{wehn_icassp03}, \cite{speziali_EUROMICRO04},
\cite{wehn_iscas05}, \cite{moussa_date07}, \cite{moussa_iscas08}. 
In addition, NOC based turbo decoder architectures have the intrinsic feature of adaptively reducing the 
communication bandwidth by the inhibition of unnecessary extrinsic information exchange. 
This can be obtained by exploiting bit-level reliability-based criteria where unnecessary iterations for reliable bits are 
avoided \cite{baghdadi_EL06}.

In \cite{wehn_icecs02}, \cite{wehn_icassp03}, \cite{speziali_EUROMICRO04} ring, chordal ring and random 
graph topologies are investigated whereas in \cite{wehn_iscas05} previous works are extended to mesh and 
toroidal topologies. Furthermore, in \cite{moussa_date07} butterfly and Benes topologies are studied, 
and in \cite{moussa_iscas08} binary de-Bruijn topologies are considered. 
However, none of these works 
presents a unified framework to design 
a NOC based turbo decoder, showing possible complexity/performance trade-offs.
This work aims at filling this gap and provides two novel contributions in the area of flexible turbo decoders: 
i) a comprehensive study of NOC based turbo decoders, conducted by means of a dedicated NOC simulator; ii) a list 
of obtained results, showing the complexity/performance trade-offs offered by different topologies, routing algorithms, 
node and ASIP architectures.

The paper is structured as follows: in section \ref{sec:system_analysis} the requirements and characteristics of a
parallel turbo decoder architecture are analyzed, whereas in section \ref{sec:noc} NOC based approach is 
introduced. Section \ref{sec:topologies} summarizes the topologies considered in previous works and introduces generalized 
de-Bruijn and generalized Kautz topologies as promising solutions for NOC based turbo decoder architectures. 
In section \ref{sec:ra} three main routing algorithms are introduced, whereas in section \ref{sec:tnoc} the Turbo NOC 
framework is described. Section \ref{sec:routing_algo_arch} describes the architecture of the different routing algorithms 
considered in this work, section \ref{sec:results} presents the experimental results and section \ref{sec:concl} draws 
some conclusions.

\section{System requirement analysis}
\label{sec:system_analysis}

A parallel turbo decoder can be modeled as $P$ processing elements that need to read from and write to $P$ memories. 
Each processing element, often referred to as soft-in-soft-out (SISO) module, performs the BCJR 
algorithm \cite{bahl_TrIT94}, whereas the memories are used for exchanging the extrinsic information $\lambda$ among 
the SISOs. 
The decoding process is iterative and usually each SISO performs sequentially the BCJR algorithm for the two constituent 
codes used at the encoder side; for further details on the SISO module the reader can refer to \cite{benedetto_ETR98}.
As a consequence, each iteration is made of two half iterations referred to as 
interleaving and de-interleaving. During one half iteration the extrinsic information produced by SISO $i$ at time  
$j$ ($\lambda_{i,j}$) 
is sent to the memory $k$ at the location $t$, where $k=k(i,j)$ and $t=t(i,j)$ are functions of $i$ and $j$
derived from 
the permutation law ($\Pi$ or interleaver) employed at the encoder side. 
Thus, the time required to complete the decoding is directly related to the number of clock cycles necessary to 
complete a half iteration. 
Without loss of generality, we can express the number of cycles required to complete a 
half iteration ($hi$) as 
\begin{equation}
N^{hi}_{cyc} = \frac{N}{P \cdot R} + IL
\end{equation}
where $N$ is the total number of trellis steps in a data frame, $N/P$ is 
the number of trellis steps processed by each SISO, 
$R$ is the SISO output rate, namely the number of trellis steps processed by 
a SISO in a clock cycle, and $IL$ is the 
interconnection structure latency.
Thus, the decoder throughput expressed as the number of decoded bits over the time required to complete the decoding 
process is
\begin{equation}
T = \frac{d\cdot N \cdot f_{clk}}{2I \cdot N^{hi}_{cyc}} = \frac{d\cdot N\cdot f_{clk}}{2I\cdot\left(\frac{N}{P\cdot R} + IL \right)}
\label{eq:T}
\end{equation}
where $f_{clk}$ is the clock frequency, $I$ is the number of iterations, $d=1$ for binary codes and $d=2$ for double 
binary codes. When the interconnection structure latency is negligible with respect to the number of cycles required 
by the SISO, we obtain
\begin{equation}
T \approx \frac{d \cdot P \cdot R}{2I} \cdot f_{clk}
\label{eq:Tapprox}
\end{equation}
\begin{figure*}[th!]
  \centering
  \includegraphics[width=\textwidth]{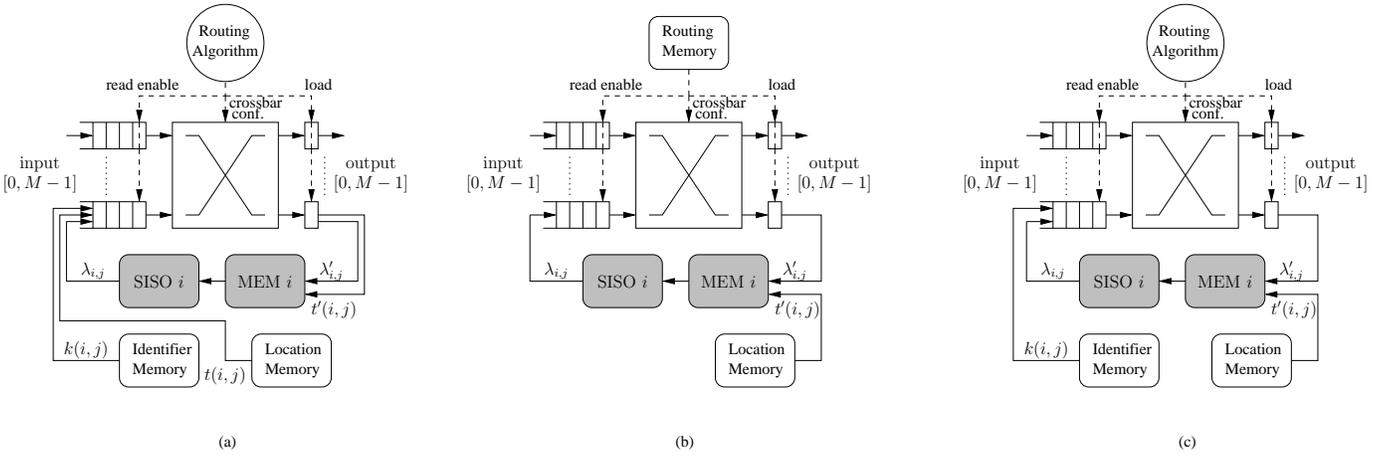}
\caption{Node block scheme: 
(a) destination identifier and memory location are sent over the network;
(b) routing algorithm is precalculated and stored in a routing memory;
(c) hybrid solution}
\label{fig:node}
\end{figure*}
Thus, to achieve a target throughput $\hat{T}$ and satisfactory error rate performance, a proper number $\hat{I}$ of 
iterations should be used. 
The minimum $P$ ($P_m$) to satisfy $\hat{T}$ with $\hat{I}$ iterations can be estimated 
from (\ref{eq:Tapprox}) for some ASIP architectures available in the literature. 
If we consider $\hat{I}=5$, as in \cite{wehn_TVLSI08}, \cite{baghdadi_TVLSI09},
$P$ ranges in [5, 37] to achieve $\hat{T}=200$ Mb/s (see Table \ref{tab:PASIP}).
It is worth pointing out that the $C=(R\cdot d)^{-1}$ values in Table \ref{tab:PASIP} 
represent the average numbers of cycles required by the SISO to update the soft information of one bit 
(see Table VI in \cite{wehn_TVLSI08} and Table I in \cite{baghdadi_TVLSI09}).
Moreover, $C$ strongly depends on the internal architecture of the SISO and in general tends to 
increase with the code complexity.
As a consequence, several conditions can further increase $P$, namely 
1) interconnection structures with larger $IL$; 
2) higher $(R \cdot d)^{-1}$ values; 
3) higher $\hat{T}$; 
4) higher $\hat{I}$; 
5) lower clock frequency.
Thus, we consider as relevant for investigation a slightly wider range for $P$: $P \in \{8, 16, 32, 64\}$.
\begin{table}[h!]
  \centering
  \caption{Parallelism degree required to obtain $\hat{T}=200$ Mb/s for $\hat{I}=5$ with some ASIP architectures available in 
the literature}
  \label{tab:PASIP}
  \begin{tabular}{|c|c|c|c|c|c|c|}
    \hline
     Architecture       & Technology  & $f_{clk}$ & $C=(R\cdot d)^{-1}$ & $P_m$ \\
                        &  [nm]  & [MHz]     &        &       \\
    \hline 
\cite{wehn_TVLSI08}     & 65     & 400       & 2.35   & 6  \\
\cite{baghdadi_TVLSI09} & 90     & 400       & 1.75   & 5  \\
\cite{baghdadi_date06}  & 90     & 335       & 6.5    & 20 \\
\cite{baghdadi_date06}  & 180    & 180       & 6.5    & 37 \\
    \hline
  \end{tabular}
\end{table}

\section{Network based approach}
\label{sec:noc}

The NOC approach \cite{beniniNOC} has been proposed as a general methodology to interconnect heterogeneous 
intellectual properties (IP) in complex systems on chip (inter-IP interconnection). 
Recent works deal with methodologies to design application specific NOCs (e.g. \cite{benini_date06}) 
where the NOC is tailored around a particular application or group of applications. 
In this scenario, turbo decoder architectures are a common IP required in physical layer chips for modern communication 
standards. 
In this work, as in some previous papers, e.g. \cite{wehn_icecs02}, \cite{wehn_iscas05}, \cite{moussa_iscas08} 
we concentrate on the problem of interconnecting 
the main building blocks of a parallel 
turbo decoder, namely 
we focus on the intra-IP interconnection problem \cite{vacca_DSD09}, 
and we do not deal with the general problem of connecting the 
turbo decoder IP to other receiver modules through an inter-IP interconnection network. 
To that purpose, it is worth pointing out that statistical characterization of 
communication patterns, which is one of the most relevant aspects in the design of application specific NOCs, 
is not required in turbo decoders, as communication patterns depend on $\Pi$.
As a consequence, given a set turbo codes with the corresponding $\Pi$ laws, the intra-IP communication patterns are 
deterministic. Thus, the challenge of NOC based turbo decoder architectures is to find one or more sets of parameters that
match throughput constraints for all supported standards 
with a reduced complexity overhead. This set of parameters includes 
$R$, $P$, the topology and the routing algorithm. 

A NOC based turbo decoder architecture relies on $P$ nodes connected through a proper topology where the 
extrinsic information is sent over the network according to a certain routing algorithm. We assume that 
each node has a certain number of input and output ports ($M$), 
a FIFO for each input, 
a crossbar to connect each input FIFO to a proper output 
and an output register, as shown in Fig. \ref{fig:node}. 
Furthermore, each node has a local SISO (SISO $i$) that sends extrinsic information over the network through 
the $M-1$ labeled input port and a local memory (MEM $i$) that receives extrinsic information from the network 
through the $M-1$ labeled output port. 

Three possible node architectures, shown in Fig. \ref{fig:node}, can be conceived to implement the node. 
\paragraph{First node architecture}
In each half iteration a SISO sends $N/P$ messages where every message is made of a payload containing 
the extrinsic information and the location of the memory where the extrinsic information will be written ($t(i,j)$), and 
a header containing the identifier of the destination node ($k(i,j)$). 
As a consequence, the node should contain a memory to store $k(i,j)$ (Identifier Memory),
a memory to store $t(i,j)$ (Location Memory) and a routing algorithm to properly route messages through 
the network (see Fig. \ref{fig:node} (a)).
\paragraph{Second node architecture}
Since the permutation law defined by the interleaver is known a-priori, the path followed by 
a message during an interleaving (or de-interleaving) half iteration can be precalculated and stored 
as a routing information into a routing memory for each node. This approach 
reduces the data width of FIFOs, crossbars and registers as neither $k(i,j)$ nor $t(i,j)$ are sent over the 
network. The location where received messages ($\lambda'_{i,j}$) will be stored ($t'(i,j)$) 
can be also precalculated and stored into a Location Memory (see Fig. \ref{fig:node} (b)). 
\paragraph{Third node architecture}
Since the routing memory foot-print can be relevant, a hybrid solution is obtained by precalculating and storing 
only $t'(i,j)$, 
whereas the routing is managed by a routing algorithm (see Fig. \ref{fig:node} (c)).
This solution does not require a Routing Memory and employs a smaller payload with respect to the solution depicted 
in Fig. \ref{fig:node} (a). On the other hand, the first approach (Fig. \ref{fig:node} (a)) directly supports adaptive 
bandwidth reduction techniques, whereas, neither the second nor the third (Fig. \ref{fig:node} (b) and (c)) do.

\begin{figure}
  \centering
  \includegraphics[width=\columnwidth]{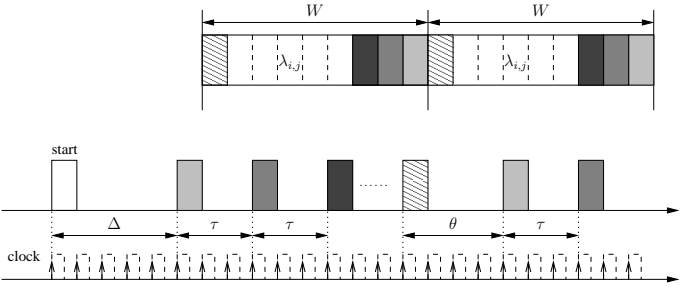}
\caption{SISO architecture parameters: graphical representation of the timing for a generic SISO architecture 
that sends the extrinsic information according to the backward recursion order}
\label{fig:SISO_params}
\end{figure}

\section{Topologies}
\label{sec:topologies}

As highlighted in \cite{wehn_iscas05} and \cite{moussa_iscas08} the choice of topologies and routing algorithms 
impacts both on throughput and complexity. As a consequence, given a certain parallelism degree $P$, topologies 
with a small node degree ($D=M-1$) as rings ($D=2$) keep the network complexity overhead limited. 
On the other hand, topologies with a higher node degree 
as toroidal meshes ($D=4$) can increase the throughput. Since interleavers tend to spread the extrinsic information 
almost uniformly among the $P$ memories, we consider fixed degree topologies where every node has the same degree $D$. 
Among fixed degree topologies we included rings and toroidal mesh networks as in \cite{wehn_icecs02} and 
\cite{wehn_iscas05}. 
Moreover, since de-Bruijn topologies have logarithmic diameter, they are good candidates to 
reduce the latency of the network in turbo decoder architectures, as investigated in \cite{moussa_iscas08}.
A de-Bruijn topology is made of nodes labeled by an array of $n$ elements, each element is taken from an 
alphabet $\mathcal{A}$ with $m$ symbols. 
Each node is connected to the nodes whose labels are obtained by left-shifting the 
node-label array and by placing in the rightmost position a symbol from $\mathcal{A}$. As a consequence, each node 
is connected to $m$ nodes ($D=m$) and the number of nodes in the network is $P=m^n$. Thus, in general, de-Bruijn 
topologies for given $P$ and $D$ values not always exist. This limitation can be overcome by using generalized 
de-Bruijn topologies \cite{imase_TC81}. 
A further limitation of de-Bruijn and generalized de-Bruijn topologies are 
self loops that are present in some nodes (e.g. the node with label zero). 
\begin{figure*}[t!]
  \centering
  \includegraphics[width=\textwidth]{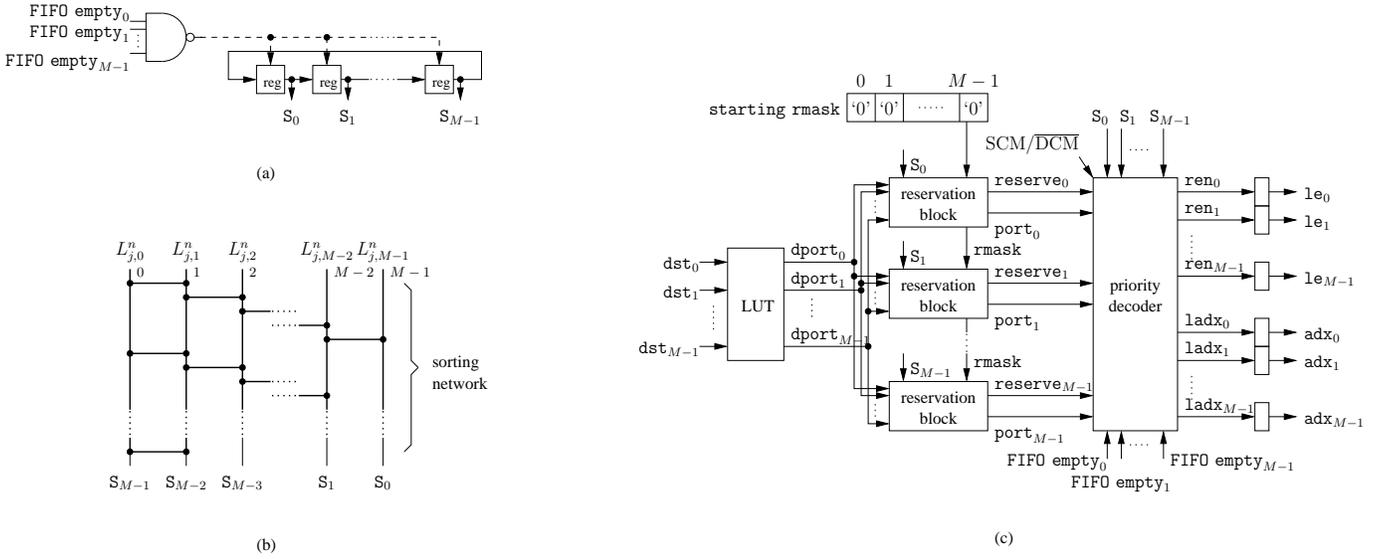}
\caption{Routing algorithm architecture: RR block scheme (a), FL block scheme (b), SSP block scheme (c)}
\label{fig:routing_algo}
\end{figure*}

This limitation is overcome by Kautz topologies where nodes are labeled as in de-Bruijn topologies but avoiding 
sequences with equal symbols in consecutive positions of the node-label array 
(Kautz sequences). 
Then, node connections are obtained as for de-Bruijn topologies, where the symbol placed in the rightmost position of 
the node-label array is taken from $\mathcal{A}$, subject to the constraint that the obtained node-label array 
is a Kautz sequence. As a consequence, each node is connected to $m-1$ nodes ($D=m-1$) and the number of
nodes in the network is $P=m\cdot(m-1)^{n-1}$. Thus, as for de-Bruijn topologies, Kautz topologies for assigned 
$P$ and $D$ values not always exist. 
This problem is eliminated by using generalized Kautz topologies 
\cite{imase_TC83}.

Moreover, we included in our investigation honeycomb networks that, 
as suggested in \cite{parhami_TPDS01}, are alternatives to toroidal meshes that reduce nodes degree to $D=3$.
Thus, 
we have that rings, honeycombs and toroidal meshes are represented as 
undirected graphs, whereas de-Bruijn (generalized de-Bruijn) and Kautz (generalized Kautz) correspond to 
directed graphs.

\section{Routing algorithms}
\label{sec:ra}

Since in turbo decoder architectures the achieved throughput is a key objective, we should try to deliver 
messages following the shortest available path. Furthermore the NOC must grant that all messages are delivered to 
the destination, namely dropping of messages to avoid dead-locks is not allowed as it could impair the decoder 
correction capability.
As highlighted in \cite{moussa_iscas08} shortest-path 
based routing algorithms are suited to achieve high throughput and grant message delivery. 
In the following we will consider both single-shortest-path (SSP) and all-local-shortest-path (ASP)
based routing algorithms.
In SSP algorithms only one shortest-path from each node $i$ to each node $k$ is considered, whereas ASP algorithms 
rely on the fact that in a topology two nodes $i$ and $k$ may be connected by more shortest-paths. 
At each node $i$, the actual routing choice toward node $k$ must be made by selecting one destination node 
directly connected to $i$ and belonging to a set $\mathcal{N}^{i,k}$ defined as the set of all nodes adjacent to 
$i$ and placed on a shortest path between $i$ and $k$.

Based on shortest-path routing, we tested three strategies to serve the input FIFOs, namely SSP Round-Robin (SSP-RR), 
SSP FIFO-length (SSP-FL) and ASP FIFO-length with traffic-spreading (ASP-FT). 
The SSP-RR approach is based on a circular serving policy coupled to the SSP approach.
The SSP-FL approach serves the input FIFOs based on the number of elements contained in each input FIFO: 
the longest FIFO is served first and the shortest one is served last. 
The ASP-FT approach is based on the input FIFO length serving policy, as for SSP-FL, but it is more complex and 
can be described as follows. Let's define 
$\mathcal{I}^{i,l}_j$ as the set of input ports in a node $l \in \mathcal{N}^{i,k}$ 
that can receive a message from node $i$ at time $j$. At time $j$ the number of elements contained in the 
input FIFO associated to port $p \in \mathcal{I}^{i,l}_j$ with $l \in \mathcal{N}^{i,k}$ is $L^l_{j,p}$.
According to Algorithm \ref{algo:ASP-FT}, the ASP-FT routing algorithm chooses 
$\hat{l} \in \mathcal{N}^{i,k}$ and $\hat{p} \in \mathcal{I}^{i,l}_j$ 
so that 
\begin{equation}
L^{\hat{l}}_{j,\hat{p}} = L_{min} = \min_{p,l}\{L^l_{j,p}\}
\label{eq:Lmin}
\end{equation}
The couples $\hat{l}$, $\hat{p}$ that satisfy (\ref{eq:Lmin}) belong to the set 
$\mathcal{J}^{i,\hat{l}}_{j,\hat{p}}$.
To choose only one couple in $\mathcal{J}^{i,\hat{l}}_{j,\hat{p}}$ 
we operate a traffic spreading based selection, namely our objective is to spread the traffic as much as 
possible over the network. To that purpose we use a set of counters ($Q$), where each counter 
$Q^{i,\hat{l}}_{j,\hat{p}}$ is incremented 
each time a message is sent from node $i$ to node $\hat{l}$ through input port $\hat{p}$.
Then, we select the couple $\tilde{l}$, $\tilde{p} \in \mathcal{J}^{i,\hat{l}}_{j,\hat{p}}$ that is associated 
to the least used path 
\begin{equation}
Q^{i,\tilde{l}}_{j,\tilde{p}}=Q_{min}=\min_{\hat{p},\hat{l}}\{Q^{i,\hat{l}}_{j,\hat{p}}\}
\end{equation}
\begin{algorithm}
  \caption{ASP-FT routing algorithm}
  \label{algo:ASP-FT}
  \begin{algorithmic}[1]
    \REQUIRE $Q^{i,l}_{j,p} \leftarrow Q^{i,l}_{j-1,p}$ and $Q^{i,l}_{0,p} \leftarrow 0$ 
    \STATE $L_{min} \leftarrow \infty$ 
    \STATE $Q_{min} \leftarrow \infty$ 
    \FORALL{$l \in \mathcal{N}^{i,k}$}
    \STATE build $\mathcal{I}^{i,l}_j$
    \FORALL{$p \in \mathcal{I}^{i,l}_j$}
    \STATE get $L^l_{j,p}$
    \IF{$L^l_{j,p} \le L_{min}$}
    \IF{$Q^{i,l}_{j,p} < Q_{min}$}
    \STATE $Q_{min} \leftarrow Q^{i,l}_{j,p}$
    \STATE $L_{min} \leftarrow L^l_{j,p}$
    \STATE $\tilde{l} \leftarrow l$
    \STATE $\tilde{p} \leftarrow p$
    \ENDIF
    \ENDIF
    \ENDFOR
    \ENDFOR
    \STATE $Q^{i,\tilde{l}}_{j,\tilde{p}}=Q^{i,\tilde{l}}_{j,\tilde{p}}+1$
  \end{algorithmic}
\end{algorithm}
\begin{figure*}[th!]
  \centering
  \includegraphics[width=\textwidth]{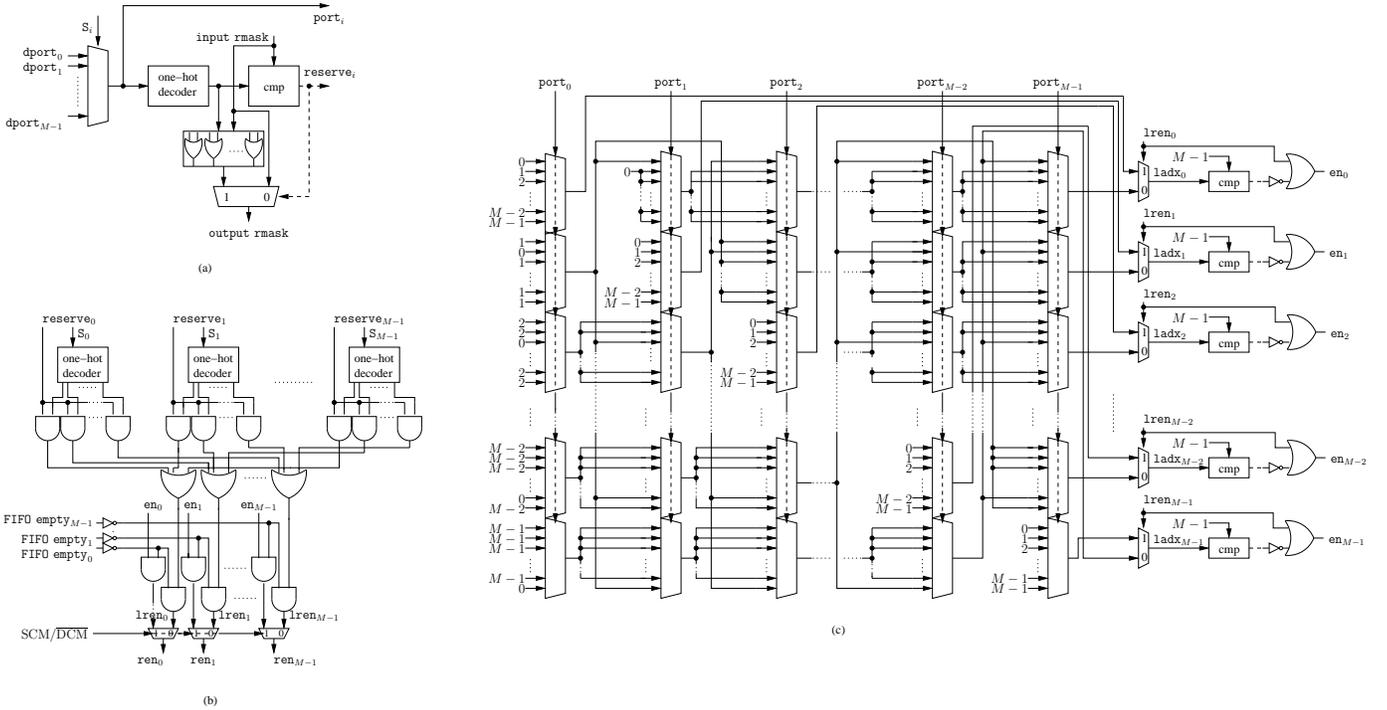}
\caption{Routing algorithm architecture details: reservation block (a), read-enable generation block (b), 
destination-port generation block (c)}
\label{fig:routing_algo_detail}
\end{figure*}

It is worth pointing out that, shortest-path based routing algorithms do not prevent output ports contention,  
that is a situation where 
two or more inputs need to send data to the same output port. 
Said $\mathcal{I}^n_{j,b}$ the set of inputs in node $n$ that at time $j$ need to send data to output port $b$, 
the contention problem can be faced by properly choosing 
an input $a \in \mathcal{I}^n_{j,b}$ allowed to send its data to output port $b$.
The remaining inputs belonging to $\mathcal{I}^n_{j,b}-\{a\}$ can be managed in different ways.
In this work we consider the following two approaches: 
i) storing $a' \in \mathcal{I}^n_{j,b}-\{a\}$ into the corresponding input FIFO so that we delay a colliding message, 
in the following we will refer to this approach as 
delay-colliding-message (DCM);
ii) if possible, sending $a' \in \mathcal{I}^n_{j,b}-\{a\}$ to 
another output port $b' \ne b$, 
send-colliding-message (SCM).
The DCM approach aims at reducing the number of hops 
to deliver a message to its destination, whereas the SCM approach aims at reducing the maximum depth of the input FIFOs.
\begin{figure}[th!]
  \centering
  \includegraphics[width=\columnwidth]{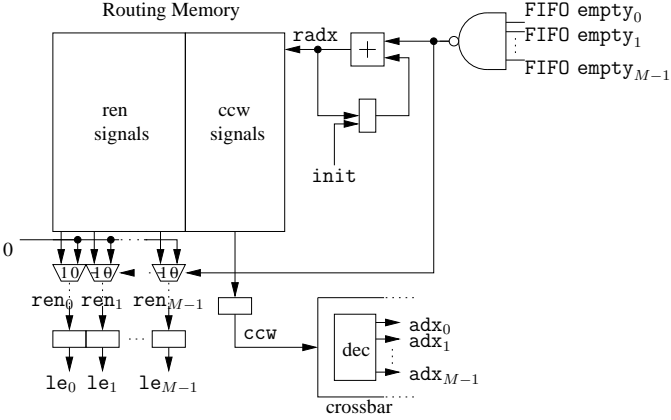}
\caption{Routing memory architecture}
\label{fig:routing_mem}
\end{figure}

\section{Turbo NOC simulator}
\label{sec:tnoc}

The Turbo NOC simulator \cite{turbo_NOC_download} is a cycle accurate, SystemC \cite{SystemC} based NOC simulator, 
specifically tailored 
for turbo decoder architectures. It estimates the throughput and complexity of a parallel NOC based turbo decoder 
architecture. 
It requires as inputs the following elements: 
the topology description in the form of an adjacence matrix, the permutation law 
used at the encoder and represented as a sequence of integer values, the routing algorithm 
(SSP-RR, SSP-FL, ASP-FT), the selected approach to handle contention (DCM/SCM) 
and the description of the key SISO characteristics. 
The required parameters to describe the SISO architecture, summarized in Fig. \ref{fig:SISO_params} are:  
\begin{enumerate}
\item the window size ($W$) \cite{benedetto_EL96},
\item the SISO latency ($\Delta$) expressed in clock cycles; $\Delta$ 
depends on the forward and backward recursion scheduling
\cite{parhi_ISCAS04}, on the trellis initialization strategy \cite{yao_ICASSP03} and on the parallelism 
level of the SISO architecture \cite{muller_ICCTA06},
\item the order used to send extrinsic informations on the network, namely forward or backward recursion 
order,
\item the number of clock cycles between two consecutive outputs $\lambda$ within a window ($\tau$), 
\item the number of clock cycles between the last output $\lambda$ of a window and the first $\lambda$ 
of the successive window ($\theta$).
\end{enumerate}
The simulator acts in two phases, a static phase (instantiation and binding) and a dynamic phase 
(cycle accurate simulation).
During the static phase, the topology description defines $P$, $D$ and all possible paths from one node to the other. 
The simulator represents the topology as a graph, calculates all the local shortest paths repeating the Floyd-Warshall 
algorithm on pruned versions of the graph until no more local paths exist between a source node $i$ and its adjacent 
nodes, and stores each result of the Floyd-Warshall algorithm as an array.
Then, if a SSP routing algorithm is employed, only the first shortest path array is employed, otherwise all the shortest 
paths are considered. 
Moreover, $P$ nodes are instantiated and binded according to the assigned topology and each SISO memory is loaded 
with $N/P$ messages, based on the assigned permutation. 
The actual decoding process executed by SISO elements is not included in the tool, which only simulates the exchange of 
extrinsic informations. However, 
the SISO architecture parameters are employed to initialize a set of counters that are used to send the extrinsic 
information over the network with the same timing as in the real SISO architecture. 
The node is described by means of 
a hardware-description-language-like (hdl-like) model. 
When the static phase is completed, the simulation starts resorting to the SystemC kernel simulator and  
performs a cycle accurate hdl-like simulation.
The results provided by the Turbo NOC simulator can be divided in two categories: 
cycle by cycle results and global results.
The cycle by cycle results are: i) for each node, the status of each FIFO, ii) for each node 
the FIFO read enable and the crossbar configuration signals, iii) for each SISO the $t'(i,j)$ sequence. 
The global results are: i) for each FIFO in each node the maximum FIFO size, ii) for each node the minimum, maximum 
and average latency (in clock cycles) of each received message and the total number of clock cycles to deliver all the 
messages.

\section{Routing algorithm architecture}
\label{sec:routing_algo_arch}

In order to keep the NOC complexity as small as possible, SSP-RR and SSP-FL routing algorithms have been 
implemented with architectures (a) and (c) in Fig. \ref{fig:node}, whereas the ASP-FT algorithm has been implemented 
as a routing memory, as in Fig. \ref{fig:node} (b). 

\subsection{SSP-RR and SSP-FL architectures}
\label{subsec:SSP}
The SSP-RR and SSP-FL architectures, thoroughly shown in Fig. \ref{fig:routing_algo} and \ref{fig:routing_algo_detail}, 
are made of two main parts. 
The first part sorts the input FIFOs based on the selected priority method (round-robin or FIFO length) and 
generates $M$ signals, $\mathtt{S}_0$, \dots, $\mathtt{S}_{M-1}$, where 
$\mathtt{S}_0$ is the label of the input that is served first and 
$\mathtt{S}_{M-1}$ is the label of the input that is served last. 
The second part serves the input FIFOs according to the order specified by the 
$\mathtt{S}_0$, \dots, $\mathtt{S}_{M-1}$  sequence and 
generates the read-enable ($\mathtt{ren}_i$) signals for the FIFOs, 
the load enable ($\mathtt{le}_i$) signals for the output registers and 
the configuration commands for the crossbar ($\mathtt{adx}_i$), 
where  $\mathtt{adx}_i$ represents the label of the destination node specified 
by the first message in FIFO $i$.
As shown in Fig. \ref{fig:routing_algo} (a), 
in the SSP-RR architecture a rotate register generates the $\mathtt{S}_i$ 
signals. On the other hand, (see Fig. \ref{fig:routing_algo} (b)), in the SSP-FL architecture the $\mathtt{S}_i$ 
signals are obtained with a sorting network. In node $n$ the sorting network takes as an input the number of elements 
contained at time $j$ in each input FIFO ($L^n_{j,p}$ with $p \in \{0, \dots, M-1\}$) and outputs 
$\mathtt{S}_0 = \arg \left\{ \max_p \{L^n_{j,p}\} \right\}$, \dots, 
$\mathtt{S}_{M-1} = \arg \left\{ \min_p \{L^n_{j,p}\} \right\}$. Both SSP-RR and SSP-FL architectures have been 
designed as parametric units to support different $M$ values.

The generation of the $\mathtt{ren}_i$, $\mathtt{adx}_i$ and $\mathtt{le}_i$ signals is enabled by the 
$\mathtt{FIFO~empty}$ signals of the input FIFOs and requires the following units:  
a look-up-table (LUT), $M$ reservation blocks and a priority decoder (Fig. \ref{fig:routing_algo} (c)). 
The LUT contains the shortest-path information. 

In the SSP approach for each node $i$ the $\mathcal{N}^{i,k}$ set contains only one node, and  
$\mathcal{I}^{i,l}_j$ contains only one port. There is only an output port on node $i$ that connects 
node $i$ with node $k$ on a shortest path. Thus, every LUT contains $P$ locations and the LUT in node $i$ 
at location $k$ contains the label of the output port to connect node $i$ to node $k$. 
As a consequence, each LUT is a $P \times \lceil \log_2(M) \rceil$ table that converts $M$ destinations 
($\mathtt{dst}_i$) to the corresponding ports ($\mathtt{dport}_i$).

The reservation blocks update an $M$-position binary mask to avoid collisions on output ports, whereas the 
priority decoder implements the selected priority and FIFO management policies by properly generating the 
the $\mathtt{ren}_i$ and $\mathtt{ladx}_i$ signals. 
Since the $\mathtt{le}_i$ and $\mathtt{adx}_i$ signals 
must be asserted the clock cycle after the $\mathtt{ren}_i$ and $\mathtt{ladx}_i$, they are delayed by means of registers. 
In particular, the $\mathtt{le}_i$ and $\mathtt{adx}_i$ are obtained by delaying the $\mathtt{ren}_i$ and $\mathtt{ladx}_i$ of one clock cycle.

\subsubsection{Reservation block}
Each reservation block (Fig. \ref{fig:routing_algo_detail} (a)) receives the $\mathtt{dport}_i$ signals, according 
to the $\mathtt{S}_0$ \dots $\mathtt{S}_{M-1}$ sequence, generates a reservation 
signal ($\mathtt{reserve}_i$) and specifies the output port to be reserved ($\mathtt{port}_i$).
The reservation is obtained by updating the $\mathtt{rmask}$, which contains a `1' in the position of 
a reserved output port and a `0' in the position of a free output port. 
Each reservation block generates 
$\mathtt{port}_i = \mathtt{dport}_{\mathtt{S}_i}$, that is converted by a one-hot decoder into a mask with a 
`1' in position $\mathtt{port}_i$. The reservation mask is updated ($\mathtt{output~rmask}$)
by comparing this mask with the
$\mathtt{input~rmask}$: if the $\mathtt{input~rmask}$ contains a `0' in position $\mathtt{port}_i$ 
the $\mathtt{reserve}_i$ goes to `1'.

\subsubsection{Priority decoder}
The priority decoder is made of two blocks: the read-enable generation block (Fig. \ref{fig:routing_algo_detail} (b)) 
and the destination-port generation block (Fig. \ref{fig:routing_algo_detail} (c)). 

\paragraph{read-enable generation block}
The read-enable generation block is based on few logic gates that act differently depeding on the approach selected 
to manage the input FIFOs (SCM/$\overline{\mathrm{DCM}}$): i) in the DCM approach, 
$\mathtt{ren}_i = \mathtt{reserve}_{\mathtt{S}_i}$ when FIFO $i$ is not empty ($\mathtt{FIFO~empty}_i$=`0') 
is obtained by combining $\mathtt{S}_i$ one-hot representation with the corresponding $\mathtt{reserve}$ signal.
ii) in the SCM approach, 
$\mathtt{ren}_i = \mathtt{en}_i$ when FIFO $i$ is not empty ($\mathtt{FIFO~empty}_i$=`0') 
is based on $\mathtt{en}_i$ that is a set of $M$ signals produced by the destination-port generation block, 
where $\mathtt{en}_i=$`0' when $\mathtt{lren}_i=$`0' and $\mathtt{ladx}_i = M-1$, namely 
the output port with label $M-1$ is used only for messages whose destination is the node itself 
(Fig. \ref{fig:routing_algo_detail} (c)).

\paragraph{destination-port generation block}
This is an array of multiplexers, where each multiplexer in position $i,i$ implements 
$\mathtt{ladx}_i = \mathtt{port}_i$ when $\mathtt{lren}_i=$`1'. On the other hand, $\mathtt{ladx}_i$ must 
take the value of an un-reserved output port when $\mathtt{lren}_i=$`0'. 
This is obtained by means of the permutation network implemented by 
the multiplexers in position $j,i$ with $j \ne i$ whose outputs ($\mathtt{mux}_{j,i}$) are
\begin{equation}
  \mathtt{mux}_{j,0} = \left\{
  \begin{array}{ll}
    0 & \mathrm{if}~\mathtt{port}_0 = j \\
    j & \mathrm{otherwise}
  \end{array} \right.
\end{equation}
and for $i>0$ 
\begin{equation}
  \mathtt{mux}_{j,i} = \left\{ 
  \begin{array} {ll}
    \mathtt{mux}_{i,i-1} & \mathrm{if}~\mathtt{port}_i = j \\
    \mathtt{mux}_{j,k}   & \mathrm{otherwise}
  \end{array} \right. 
\end{equation}
where
\begin{equation}
k = \left\{
  \begin{array} {ll}
    i-2 & \mathrm{if}~j=i-1 \\
    i-1 & \mathrm{otherwise}
  \end{array} \right.
\end{equation}
and if $k<0$, then $\mathrm{mux}_{j,k}=0$.

\subsection{ASP-FT architecture}
\label{subsec:ASP}
The ASP-FT algorithm is simply implemented by means of a routing memory. As a consequence, 
DCM and SCM 
approaches are integrated by filling the routing memory with the appropriate configuration words.
Each word is the concatenation of the $\mathtt{ren}_0$, \dots $\mathtt{ren}_{M-1}$ signals with the 
$\mathtt{adx}_0$, \dots, $\mathtt{adx}_{M-1}$ signals. In order to reduce the word width, 
the $\mathtt{adx}_0$, \dots, $\mathtt{adx}_{M-1}$ signals, which can be represented 
on $M \times \lceil \log_2 (M) \rceil$ bits,  
are coded into a crossbar configuration word ($\mathtt{ccw}$). 
Since for an $M$-port crossbar the possible configurations are $M!$, $\mathtt{ccw}$ is 
represented on $\lceil \log_2(M!)\rceil$ bits.  
The corresponding decoder is hardwired into the crossbar. Thus, as shown in Fig. \ref{fig:routing_mem} the 
main component in the routing memory architecture is a RAM.
The RAM address ($\mathtt{radx}$) 
is generated by an adder and a register. 
The adder is incremented when at least one input FIFO is not empty ($\mathtt{FIFO~empty}_i=$`0') and it is initialized to 
zero when the half iteration starts ($\mathtt{init}$). Moreover, the $\mathtt{ren}_i$ are forced to `0' when FIFOs are 
empty.

\subsection{Architecture implementation}
To achieve high throughput the routing algorithm should be able to serve the input FIFOs in one clock cycle. 
This requirement, which is an intrinsic feature of the routing memory architecture used for the ASP-FT,  
implies that the architectures for the SSP-RR and SSP-FL routing algorithms are combinational circuits. 
As it can be inferred from Fig. \ref{fig:routing_algo} and \ref{fig:routing_algo_detail} 
the speed of SSP-RR and SSP-FL architectures depends mainly on $M$, 
in fact, $M$ impacts on the size of several parts of the routing algorithm architectures, namely 
the sorting network, the shortest-path information LUT, the reservation mask, the priority decoder and on the number 
of reservation blocks.
Given the topologies presented in section \ref{sec:topologies}, we described the SSP-RR and SSP-FL architectures 
as parametric blocks and we performed the logical synthesis on a 130 nm standard cell technology for $M \in \{3, 4, 5\}$. 
Post synthesis results confirm that a clock frequency of more than 200 MHz is achieved with a complexity that ranges 
from about 1000 $\mu$m$^2$ to about 6000 $\mu$m$^2$.
\begin{table*}[t!]
  \centering
  \caption{Throughput [Mb/s]/area [mm$^2$] achieved for the WiMax interleaver ($N$=2400) with different topologies, 
$P$, $R$ and routing algorithms (DCM approach)} 
  \label{tab:wimax_results}
   { \scriptsize
  \begin{tabular}{|c|c|c|c|c|c|c|c|c|c|}
\hline
& 	 & \multicolumn{4}{c|}{$D$=2, ring} & \multicolumn{4}{c|}{$D$=2, generalized Kautz} \\
\hline
& 	 & $P$=8 & $P$=16 & $P$=32 & $P$=64 & $P$=8 & $P$=16 & $P$=32 & $P$=64 \\
\hline
\multirow{3}{*}{$R$=1.00} & SSP-RR (c) & 115.50/1.64 & 132.30/3.85 & 147.06/5.71 & 152.28/6.91 & 104.35/2.05 & 140.85/3.48 & 195.44/5.17 & 270.27/7.17 \\
& SSP-FL (c) & 112.89/1.56 & 134.38/3.18 & 147.78/4.41 & 139.86/5.43 & 108.99/1.85 & 149.07/3.21 & 209.06/4.65 & 287.77/6.35 \\
& ASP-FT (b) & 130.15/1.40 & 144.75/2.87 & 152.87/3.97 & 142.35/5.02 & 108.99/1.73 & 149.07/2.90 & 209.06/4.07 & 287.77/5.41 \\
\hline
\multirow{3}{*}{$R$=0.50} & SSP-RR (c) & 86.15/0.43 & 122.32/1.61 & 133.48/3.96 & 137.77/5.45 & 86.21/0.44 & 131.15/1.33 & 172.91/3.19 & 229.89/5.29 \\
& SSP-FL (c) & 86.15/0.42 & 123.71/1.37 & 132.89/3.67 & 130.15/5.17 & 86.15/0.41 & 138.25/1.06 & 188.68/2.62 & 241.94/4.59 \\
& ASP-FT (b) & 86.02/0.49 & 134.53/1.29 & 137.30/3.35 & 130.72/4.80 & 86.15/0.46 & 138.25/1.05 & 188.68/2.38 & 241.94/3.99 \\
\hline
\multirow{3}{*}{$R$=0.33} & SSP-RR (c) & 57.80/0.41 & 101.10/0.74 & 122.08/2.96 & 125.92/5.01 & 57.86/0.39 & 102.48/0.75 & 155.44/1.82 & 195.76/3.97 \\
& SSP-FL (c) & 57.78/0.39 & 100.84/0.72 & 121.58/2.67 & 120.97/4.87 & 57.80/0.38 & 102.21/0.67 & 161.51/1.43 & 207.25/3.28 \\
& ASP-FT (b) & 57.83/0.46 & 100.84/0.81 & 122.70/2.52 & 121.70/4.56 & 57.80/0.44 & 102.21/0.74 & 161.51/1.40 & 207.25/2.94 \\
\hline
\hline
& 	 & \multicolumn{4}{c|}{$D$=3, honeycomb} & \multicolumn{4}{c|}{$D$=3, generalized Kautz} \\
\hline
& 	 & $P$=8 & $P$=16 & $P$=32 & $P$=64 & $P$=8 & $P$=16 & $P$=32 & $P$=64 \\
\hline
\multirow{3}{*}{$R$=1.00} & SSP-RR (c) & 113.21/1.69 & 184.05/2.29 & 181.27/5.16 & 323.45/6.47 & 156.45/0.83 & 203.74/2.06 & 314.96/3.60 & 428.57/5.84 \\
& SSP-FL (c) & 114.83/1.67 & 187.79/2.22 & 179.64/4.37 & 314.96/5.99 & 166.67/0.67 & 229.89/1.93 & 332.41/3.41 & 451.13/5.49 \\
& ASP-FT (b) & 127.93/1.51 & 247.42/1.64 & 242.91/3.62 & 385.85/4.93 & 166.67/0.67 & 250.52/1.58 & 339.94/2.98 & 456.27/4.64 \\
\hline
\multirow{3}{*}{$R$=0.50} & SSP-RR (c) & 85.96/0.45 & 151.32/0.91 & 160.64/3.52 & 267.26/4.69 & 86.52/0.45 & 152.67/0.86 & 242.91/1.81 & 331.49/3.76 \\
& SSP-FL (c) & 85.90/0.43 & 151.52/0.84 & 163.71/3.10 & 261.44/4.20 & 86.52/0.44 & 152.48/0.82 & 241.94/1.66 & 338.03/3.37 \\
& ASP-FT (b) & 85.90/0.49 & 151.32/0.90 & 213.52/2.08 & 305.34/3.48 & 86.52/0.53 & 152.28/0.88 & 244.40/1.58 & 337.08/2.98 \\
\hline
\multirow{3}{*}{$R$=0.33} & SSP-RR (c) & 57.72/0.40 & 102.48/0.80 & 144.75/2.27 & 223.88/3.49 & 58.00/0.44 & 103.18/0.78 & 167.13/1.56 & 243.90/3.15 \\
& SSP-FL (c) & 57.72/0.40 & 102.48/0.79 & 148.88/2.01 & 226.42/3.21 & 58.00/0.44 & 103.09/0.78 & 168.07/1.50 & 240.48/2.96 \\
& ASP-FT (b) & 57.75/0.46 & 102.65/0.86 & 162.38/1.58 & 233.92/2.87 & 58.00/0.50 & 103.09/0.83 & 168.07/1.48 & 242.91/2.70 \\
\hline
\hline
& 	 & \multicolumn{4}{c|}{$D$=4, toroidal mesh} & \multicolumn{4}{c|}{$D$=4, generalized Kautz} \\
\hline
& 	 & $P$=8 & $P$=16 & $P$=32 & $P$=64 & $P$=8 & $P$=16 & $P$=32 & $P$=64 \\
\hline
\multirow{3}{*}{$R$=1.00} & SSP-RR (c) & 123.84/1.17 & 171.67/2.10 & 187.21/4.24 & 310.08/6.24 & 140.19/0.94 & 268.46/1.56 & 334.26/3.31 & 517.24/5.29 \\
& SSP-FL (c) & 129.87/1.14 & 167.83/2.00 & 200.33/3.67 & 323.45/5.77 & 155.24/0.82 & 281.69/1.30 & 347.83/3.13 & 550.46/4.92 \\
& ASP-FT (b) & 165.29/0.69 & 282.35/1.30 & 357.14/2.79 & 497.93/4.52 & 167.13/0.67 & 281.69/1.24 & 397.35/2.49 & 550.46/4.28 \\
\hline
\multirow{3}{*}{$R$=0.50} & SSP-RR (c) & 86.15/0.46 & 147.78/1.03 & 165.29/2.75 & 260.30/4.58 & 86.52/0.47 & 153.65/0.89 & 248.45/1.91 & 359.28/3.64 \\
& SSP-FL (c) & 86.08/0.45 & 151.52/0.95 & 178.31/2.51 & 270.27/4.25 & 86.52/0.46 & 153.85/0.88 & 247.93/1.81 & 360.36/3.50 \\
& ASP-FT (b) & 86.21/0.54 & 152.28/1.05 & 242.42/1.85 & 334.26/3.46 & 86.52/0.57 & 153.85/0.96 & 248.96/1.77 & 360.36/3.27 \\
\hline
\multirow{3}{*}{$R$=0.33} & SSP-RR (c) & 57.80/0.44 & 102.92/0.93 & 154.64/1.97 & 223.46/3.81 & 58.00/0.46 & 103.54/0.88 & 169.01/1.75 & 248.96/3.46 \\
& SSP-FL (c) & 57.80/0.44 & 102.92/0.91 & 163.49/1.79 & 226.42/3.57 & 58.00/0.45 & 103.54/0.86 & 169.25/1.69 & 248.45/3.34 \\
& ASP-FT (b) & 57.83/0.50 & 102.92/0.98 & 166.90/1.79 & 238.57/3.25 & 58.00/0.53 & 103.54/0.92 & 169.25/1.68 & 248.45/3.15 \\
\hline
  \end{tabular}
}
\end{table*}
\begin{table*}[t!]
  \centering
  \caption{Throughput [Mb/s]/area [mm$^2$] achieved for the circular shifting interleaver ($N$=24576) with different 
topologies, $P$, $R$ and routing algorithms with DCM approach. 
Light-gray, mid-gray and dark-gray cells indicate the highest throughput, the highest area and the lowest area points 
for each $D$ value respectively} 
  \label{tab:mhoms_results}
   { \scriptsize
  \begin{tabular}{|c|c|c|c|c|c|c|c|c|c|}
\hline
& 	 & \multicolumn{4}{c|}{$D$=2, ring} & \multicolumn{4}{c|}{$D$=2, generalized Kautz} \\
\hline
& 	 & $P$=8 & $P$=16 & $P$=32 & $P$=64 & $P$=8 & $P$=16 & $P$=32 & $P$=64 \\
\hline
\multirow{3}{*}{$R$=1.00} & SSP-RR (c) & 62.56/6.58 & 72.23/15.73 & 81.22/24.45 & \cellcolor[gray]{0.8} 87.04/30.37 & 56.62/8.43 & 77.26/14.10 & 116.01/20.56 & \cellcolor[gray]{0.8} 169.96/26.72 \\
& SSP-FL (c) & 62.57/6.84 & 73.89/13.90 & 82.98/18.67 & 88.25/22.11 & 59.52/8.09 & 83.52/13.50 & 125.31/18.55 & 183.79/23.53 \\
& ASP-FT (b) & 71.48/5.93 & 81.57/11.29 & 88.17/15.12 & \cellcolor[gray]{0.9} 91.14/19.36 & 59.52/6.94 & 83.52/10.74 & 125.31/13.90 & 183.79/16.74 \\
\hline
\multirow{3}{*}{$R$=0.50} & SSP-RR (c) & 49.12/1.80 & 72.36/6.00 & 80.26/15.71 & 86.11/21.02 & 49.13/1.79 & 77.37/4.10 & 114.99/10.17 & 165.12/16.37 \\
& SSP-FL (c) & 49.12/1.78 & 73.42/5.10 & 82.28/14.68 & 87.29/20.48 & 49.13/1.78 & 86.74/2.98 & 129.59/8.00 & 186.75/13.78 \\
& ASP-FT (b) & 49.12/2.50 & 82.40/4.93 & 87.48/12.55 & 90.23/18.42 & 49.13/2.39 & 86.74/3.43 & 129.59/7.03 & \cellcolor[gray]{0.9} 186.75/10.80 \\
\hline
\multirow{3}{*}{$R$=0.33} & SSP-RR (c) & \cellcolor[gray]{0.7} 32.78/1.76 & 63.67/2.09 & 79.52/11.40 & 85.06/19.10 & 32.78/1.76 & 63.83/2.06 & 111.61/4.13 & 162.20/9.53 \\
& SSP-FL (c) & \cellcolor[gray]{0.7} 32.78/1.76 & 63.68/2.04 & 81.51/9.65 & 86.27/18.43 & \cellcolor[gray]{0.7} 32.77/1.75 & 63.81/2.01 & 123.57/2.66 & 186.58/6.51 \\
& ASP-FT (b) & 32.78/2.51 & 63.67/3.37 & 86.71/9.32 & 88.90/17.17 & 32.77/2.44 & 63.81/2.99 & 123.57/3.71 & 186.58/6.45 \\
\hline
\hline
& 	 & \multicolumn{4}{c|}{$D$=3, honeycomb} & \multicolumn{4}{c|}{$D$=3, generalized Kautz} \\
\hline
& 	 & $P$=8 & $P$=16 & $P$=32 & $P$=64 & $P$=8 & $P$=16 & $P$=32 & $P$=64 \\
\hline
\multirow{3}{*}{$R$=1.00} & SSP-RR (c) & 63.28/6.51 & 107.23/8.07 & 103.96/20.19 & \cellcolor[gray]{0.8} 219.19/21.14 & 87.61/2.86 & 118.27/6.64 & 210.05/10.39 & 332.65/16.23 \\
& SSP-FL (c) & 64.03/6.67 & 109.73/8.31 & 106.61/16.64 & 214.53/20.29 & 97.37/2.06 & 135.54/6.08 & 220.22/10.86 & \cellcolor[gray]{0.8} 350.29/16.37 \\
& ASP-FT (b) & 72.48/5.74 & 152.42/5.28 & 160.67/11.92 & \cellcolor[gray]{0.9} 313.79/13.45 & 97.37/2.29 & 153.26/4.55 & 239.53/8.07 & \cellcolor[gray]{0.9} 375.55/11.66 \\
\hline
\multirow{3}{*}{$R$=0.50} & SSP-RR (c) & 49.12/1.80 & 95.52/2.19 & 102.95/12.05 & 213.78/10.75 & 49.16/1.81 & 95.63/2.15 & 185.62/2.91 & 322.18/5.24 \\
& SSP-FL (c) & 49.12/1.79 & 95.48/2.13 & 106.06/10.62 & 208.55/9.56 & 49.16/1.81 & 95.69/2.10 & 185.68/2.76 & 346.34/4.32 \\
& ASP-FT (b) & 49.12/2.50 & 95.60/3.05 & 163.71/4.91 & 312.99/6.14 & 49.16/2.63 & 95.69/3.02 & 185.62/3.54 & 348.10/4.60 \\
\hline
\multirow{3}{*}{$R$=0.33} & SSP-RR (c) & 32.78/1.77 & 63.84/2.08 & 102.14/5.83 & 205.69/5.13 & \cellcolor[gray]{0.7} 32.79/1.79 & 63.89/2.08 & 124.30/2.68 & 235.31/3.96 \\
& SSP-FL (c) & \cellcolor[gray]{0.7} 32.78/1.76 & 63.84/2.06 & 108.17/4.43 & 216.03/4.54 & \cellcolor[gray]{0.7} 32.79/1.79 & 63.89/2.07 & 124.27/2.62 & 235.58/3.78 \\
& ASP-FT (b) & 32.78/2.50 & 63.85/2.99 & 123.62/4.18 & 233.35/5.14 & 32.79/2.47 & 63.89/2.95 & 124.30/3.62 & 235.58/4.68 \\
\hline
\hline
& 	 & \multicolumn{4}{c|}{$D$=4, toroidal mesh} & \multicolumn{4}{c|}{$D$=4, generalized Kautz} \\
\hline
& 	 & $P$=8 & $P$=16 & $P$=32 & $P$=64 & $P$=8 & $P$=16 & $P$=32 & $P$=64 \\
\hline
\multirow{3}{*}{$R$=1.00} & SSP-RR (c) & 70.18/4.23 & 97.20/6.89 & 110.05/13.77 & \cellcolor[gray]{0.8} 202.77/16.59 & 77.99/3.34 & 174.15/3.71 & 215.50/8.33 & \cellcolor[gray]{0.8} 493.89/10.22 \\
& SSP-FL (c) & 73.67/4.04 & 96.02/6.58 & 117.84/11.43 & 214.68/15.00 & 86.09/3.21 & 184.17/2.97 & 232.38/8.38 & 516.74/9.82 \\
& ASP-FT (b) & 96.57/2.36 & 184.12/3.13 & 275.76/6.43 & \cellcolor[gray]{0.9} 471.89/9.19 & 97.11/2.35 & 184.17/3.06 & 298.83/5.09 & \cellcolor[gray]{0.9} 516.74/7.61 \\
\hline
\multirow{3}{*}{$R$=0.50} & SSP-RR (c) & 49.14/1.82 & 95.26/2.32 & 109.34/7.13 & 198.32/7.77 & 49.16/1.82 & 95.75/2.19 & 185.62/2.95 & 350.48/4.47 \\
& SSP-FL (c) &  49.14/1.80 & 95.48/2.22 & 119.74/6.10 & 213.85/7.03 & 49.16/1.82 & 95.75/2.17 & 185.96/2.89 & 350.89/4.28 \\
& ASP-FT (b) & 49.14/2.66 & 95.61/3.34 & 185.17/4.03 & 347.12/5.17 & 49.16/2.72 & 95.75/3.15 & 185.90/3.83 & 350.89/4.96 \\
\hline
\multirow{3}{*}{$R$=0.33} & SSP-RR (c) & 32.78/1.79 & 63.85/2.17 & 110.01/3.59 & 196.55/5.02 & 32.79/1.81 & 63.91/2.15 & 124.37/2.82 & 236.04/4.17 \\
& SSP-FL (c) & \cellcolor[gray]{0.7} 32.78/1.78 & 63.85/2.15 & 123.10/2.97 & 216.80/4.56 & \cellcolor[gray]{0.8} 32.79/1.80 & 63.92/2.14 & 124.40/2.78 & 235.94/4.06 \\
& ASP-FT (b) & 32.78/2.42 & 63.85/3.03 & 124.00/3.95 & 234.15/5.26 & 32.79/2.51 & 63.92/3.02 & 124.40/3.72 & 235.94/4.96 \\
\hline
  \end{tabular}
}
\end{table*}

\section{Simulations and results}
\label{sec:results}

The Turbo NOC simulator has been used to simulate both interleaving and de-interleaving with four significant 
permutation laws, namely:
\begin{enumerate}
\item WiMax interleaver with $N$=2400 and $W$=38
\item UMTS interleaver with $N$=5114 and $W$=40
\item A prunable S-random interleaver \cite{dinoi_TCom05} with $N$=16384 and $W$=37
\item A circular shifting interleaver \cite{dolinar_TDA95} with $N$=24576 and $W$=39 
\end{enumerate}
We tested the following topologies:
\begin{enumerate}
\item ring (R)
\item toroidal mesh (T)
\item honeycomb (H)
\item generalized de-Bruijn (B)
\item generalized Kautz (K)
\end{enumerate}
for $P \in \{8, 16, 32, 64\}$, with SSP-RR, SSP-FL and ASP-FT routing algorithms and including DCM and SCM 
approaches for FIFO management. The SISO architecture parameters were set as follows:
$\Delta=W/R$, $\theta=\tau=R^{-1}$ and backward recursion sending order.
For each case, the Turbo NOC simulator provided the total number of cycles required to perform a complete iteration
(interleaving and de-interleaving), the depth of each FIFO in the network, the content of each routing memory 
(see Fig. \ref{fig:node} (b)) and the $t'(i,j)$ sequence to be stored into the location memory 
(see Fig. \ref{fig:node} (b) and (c)). 
As a consequence, for each case we can estimate the achieved throughput for a certain clock frequency with a given 
number of iterations. Moreover, to characterize the complexity of each solution we give the synthesis results of 
all simulated networks for a 130 nm 
standard cell technology. Memories have been generated by means of a 130 nm memory generator. The area results concern  
all the nodes in the network where each node includes the blocks depicted in Fig. \ref{fig:node} except the SISO 
and the memory used to store the extrinsic information (shaded gray blocks in Fig. \ref{fig:node}).
As a significant case of study we consider each extrinsic information value represented on 8 bits. 
Thus, we represented $\lambda$ on 8 bits for all the simulations, except the 
ones related to the WiMax permutation law. In fact, since the WiMax turbo code is double binary, its extrinsic information 
is an array made of three log-likelihood ratios, as a consequence a message is represented on 24 bits. 
Moreover, we consider $f_{clk}=200$ MHz and $\hat{I}=8$; thus, from (\ref{eq:Tapprox}) we can infer that to sustain 
a target throughput of $\hat{T}=200$ Mb/s, we need at least $d \cdot P \cdot R = 16$, 
namely at least $P \cdot R = 16$ 
for binary codes and 
at least $P \cdot R = 8$ for double binary codes. However, due to the $IL$ term in (\ref{eq:T}), higher values 
of the $P \cdot R$ product are also of interest. 

The analysis of the experimental results obtained with the Turbo NOC simulator shows some interesting general properties. 
\begin{enumerate}
\item SSP solutions adopting the node architecture depicted in Fig. \ref{fig:node} (a) 
are the most demanding implementations in terms of area. Since the node architecture in Fig. \ref{fig:node} (c) achieves 
the same throughput as the solution in Fig. \ref{fig:node} (a) with a lower area, in the following only the node 
architecture in Fig. \ref{fig:node} (c) will be addressed.
\item The DCM FIFO management method performs better than the SCM one both in terms of throughput and complexity. 
As a consequence, in the following only results that are referred to the DCM approach will be presented.
\item Generalized de-Bruijn and generalized Kautz topologies achieve nearly the same results 
both in terms of throughput and complexity. In the following only results obtained with generalized Kautz topologies will 
be presented.
\item Results tend to be clustered into two families, namely short interleavers (WiMax interleaver with $N$=2400 and UMTS interleaver with 
$N$=5114) and long interleavers (prunable S-random interleaver with $N$=16384 and circular shifting interleaver 
with $N$=24576). For the sake of clarity, in the following, only results obtained for the WiMax interleaver ($N$=2400) and 
circular shifting interleaver ($N$=24576) will be presented.
\end{enumerate}

The most significant experimental results are summarized in Table \ref{tab:wimax_results} and \ref{tab:mhoms_results} that
refer to the WiMax interleaver with $N=2400$ and to the circular shifting interleaver with $N=24576$ respectively. 
Each cell of the two tables gives the throughput in Mb/s and the area in mm$^2$ obtained for different 
$P$ and $R$ values, routing algorithms and architectures for the DCM approach. 
In Table \ref{tab:mhoms_results} light-gray, mid-gray and dark-gray cells indicate 
the highest throughput, the highest area and the lowest area points for each $D$ value respectively.
\begin{figure*}
  \begin{minipage}[b]{.50\linewidth}
    \centering
    \centerline{\scriptsize{(a) WiMax, $N=2400$}} 
   \centerline{
      \includegraphics[width=\linewidth]{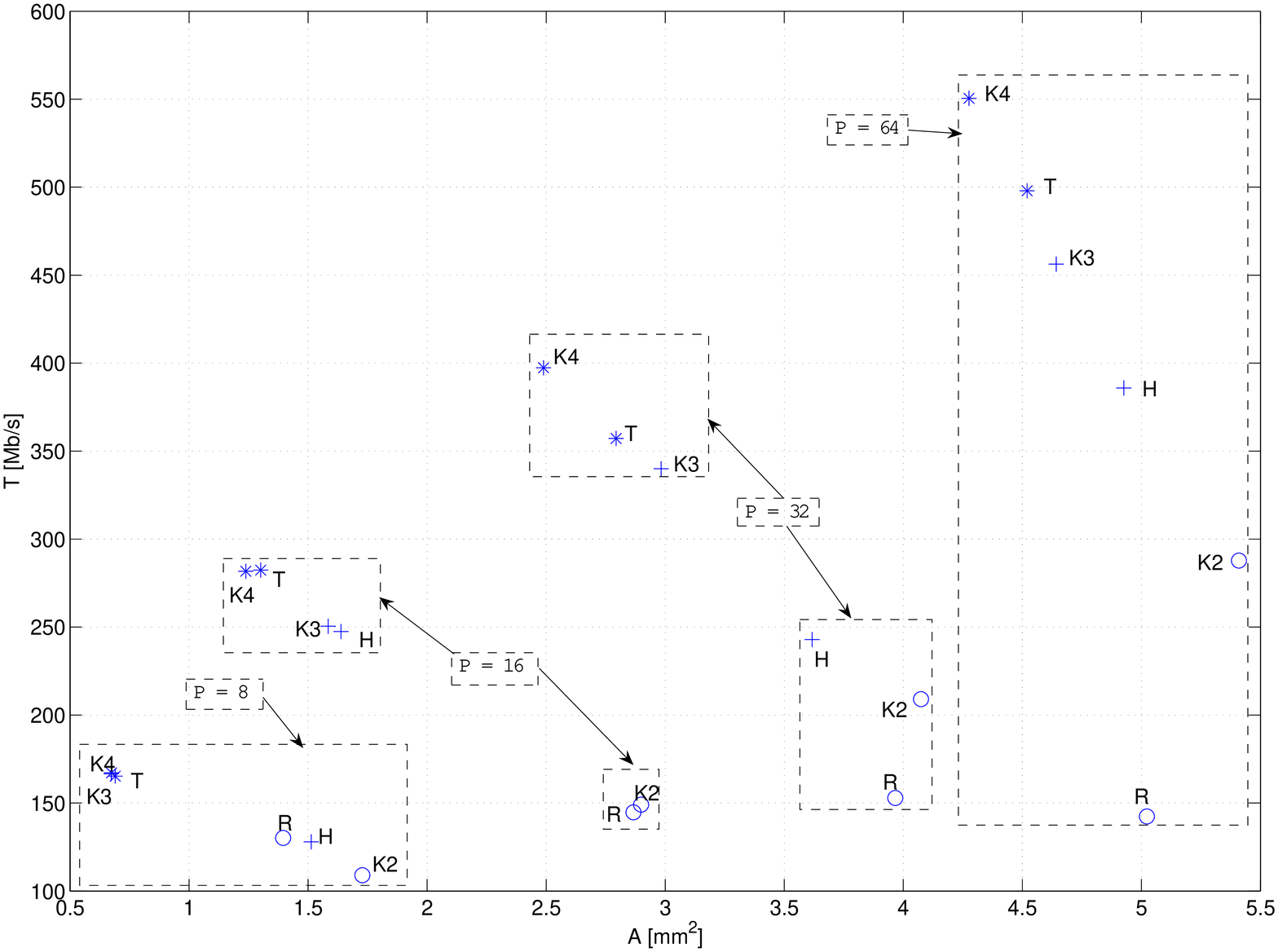}
    } \medskip
    \end{minipage}
  \hfill
  \begin{minipage}[b]{.50\linewidth}
    \centering
    \centerline{\scriptsize{(b) circular shifting interleaver, $N=24576$}} 
    \centerline{
      \includegraphics[width=\linewidth]{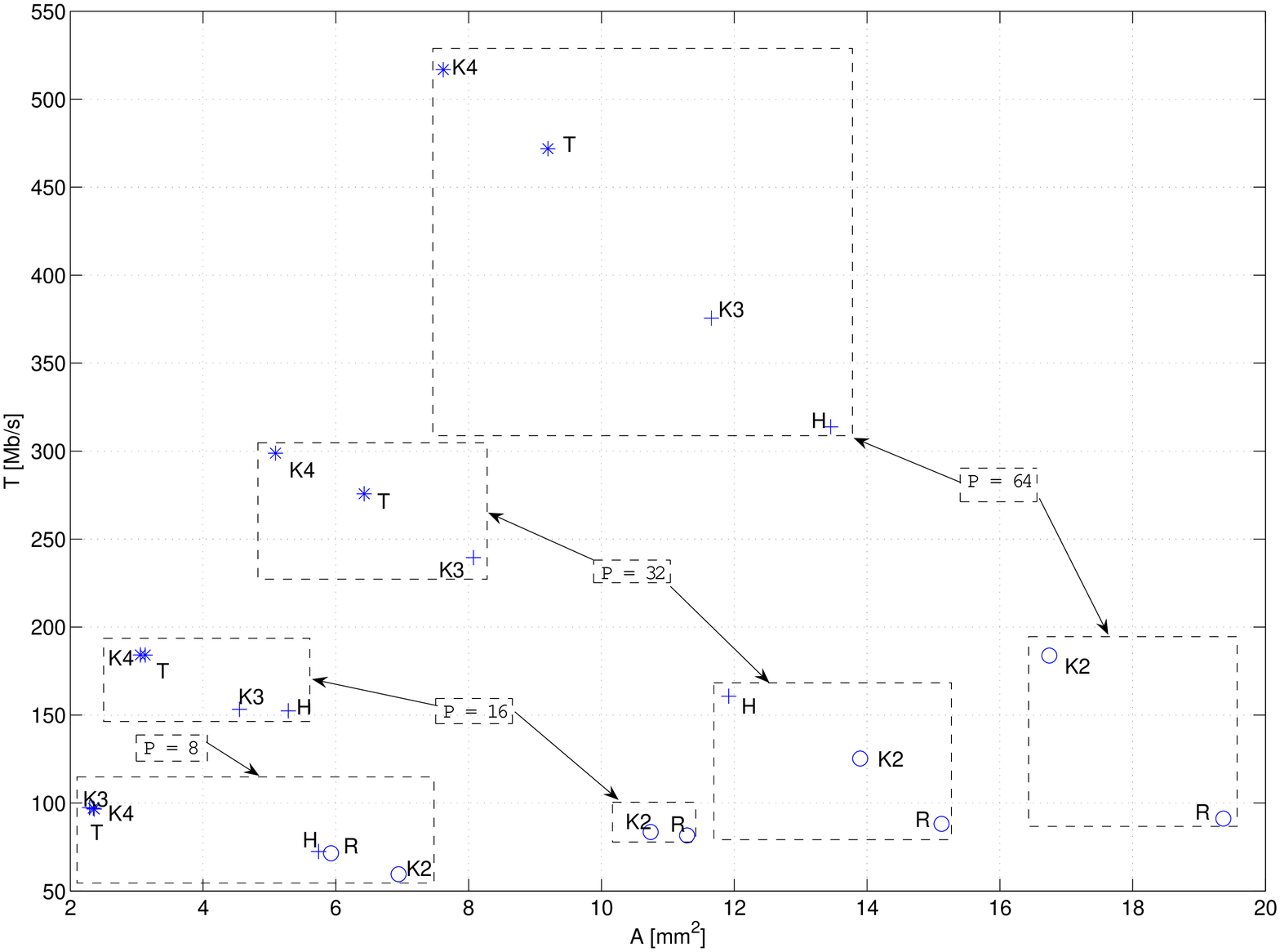}
    } \medskip
  \end{minipage}
\caption{Throughput/area comparison of different topologies for the case $R=1$, ASP-FT routing algorithm, DCM 
approach} 
\label{fig:R1_ASP-FT}
\end{figure*}
\begin{figure*}
  \begin{minipage}[b]{.50\linewidth}
    \centering
    \centerline{\scriptsize{(a) WiMax, $N=2400$, $R=1$, $P=64$}} 
   \centerline{
      \includegraphics[width=\linewidth]{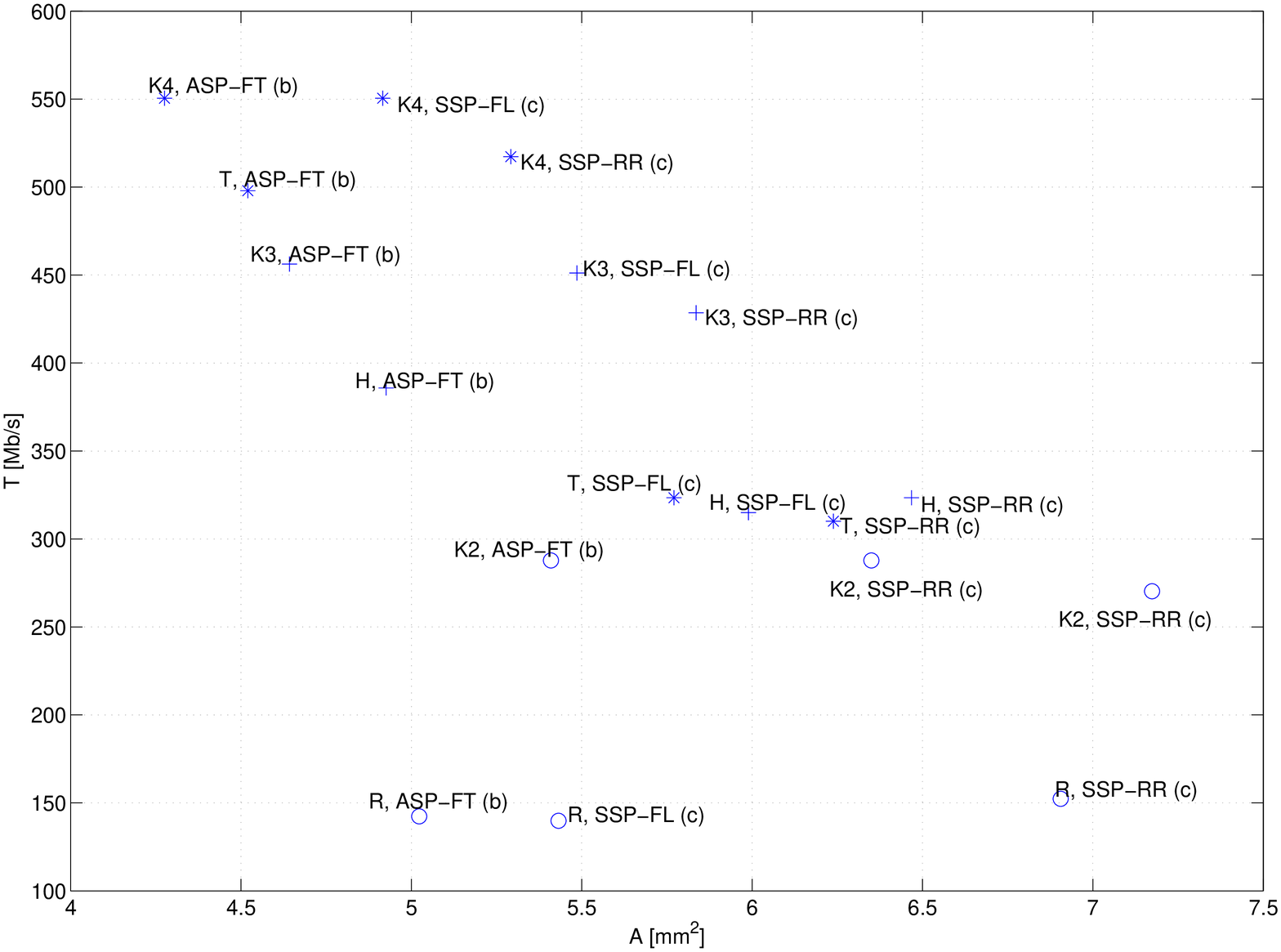}
    } \medskip
    \end{minipage}
  \hfill
  \begin{minipage}[b]{.50\linewidth}
    \centering
    \centerline{\scriptsize{(b) WiMax, $N=2400$, $R=0.33$, $P=16$}} 
    \centerline{
      \includegraphics[width=\linewidth]{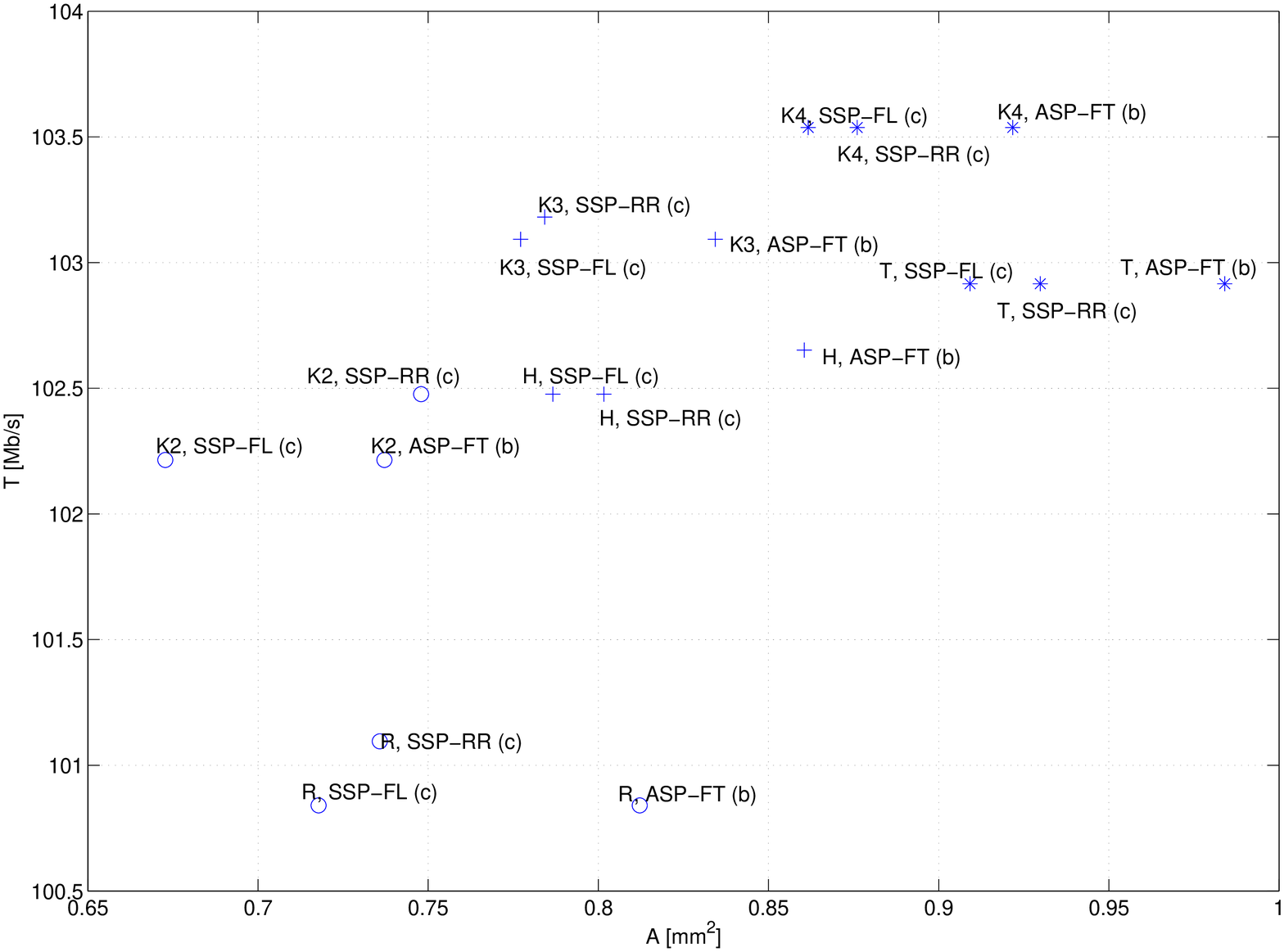}
    } \medskip
  \end{minipage}
  \begin{minipage}[b]{.50\linewidth}
    \centering
    \centerline{\scriptsize{(c) circular shifting interleaver, $N=24576$, $R=1$, $P=64$}} 
   \centerline{
      \includegraphics[width=\linewidth]{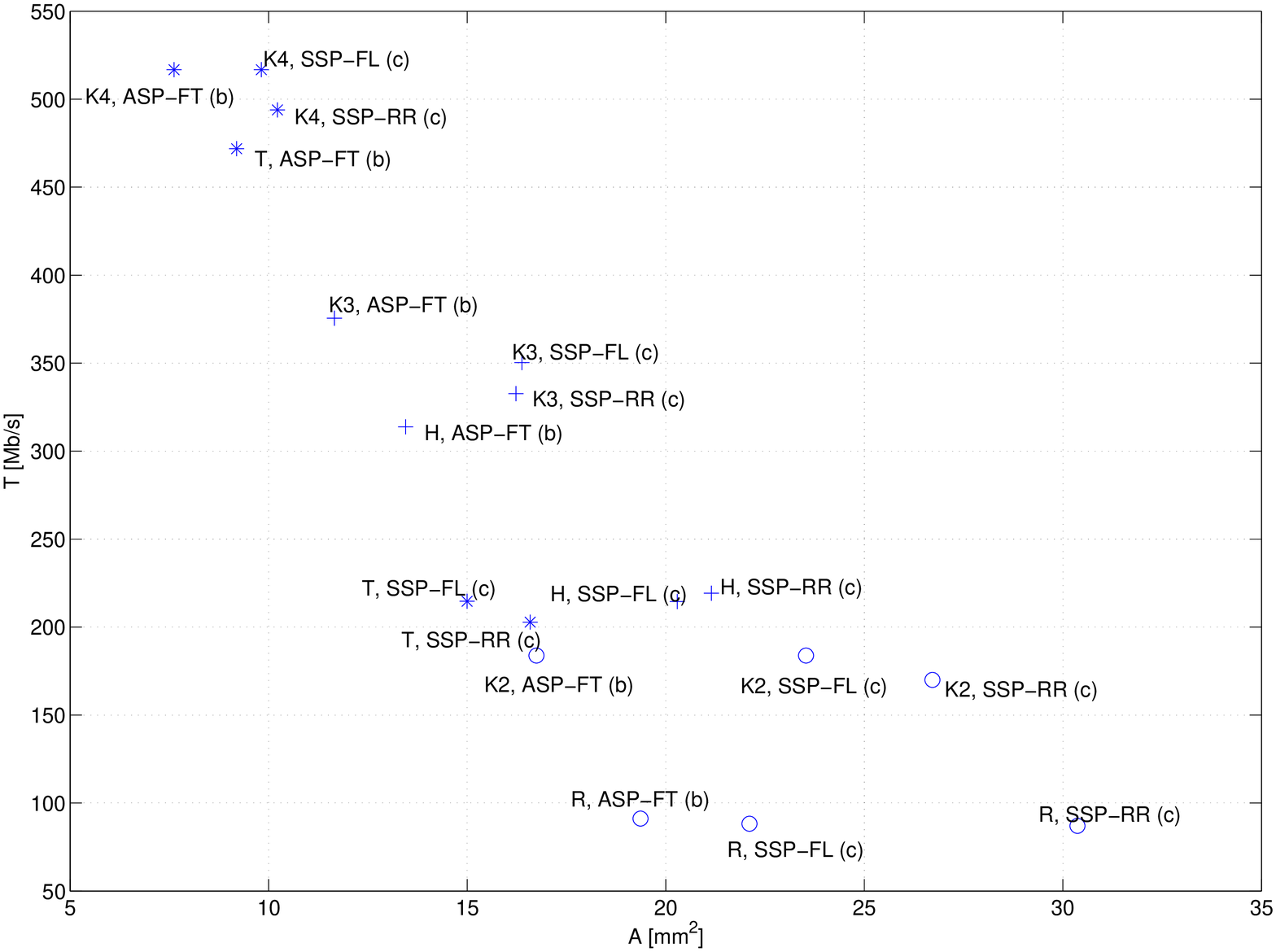}
    } \medskip
    \end{minipage}
  \hfill
  \begin{minipage}[b]{.50\linewidth}
    \centering
    \centerline{\scriptsize{(d) circular shifting interleaver, $N=24576$, $R=0.33$, $P=16$}} 
    \centerline{
      \includegraphics[width=\linewidth]{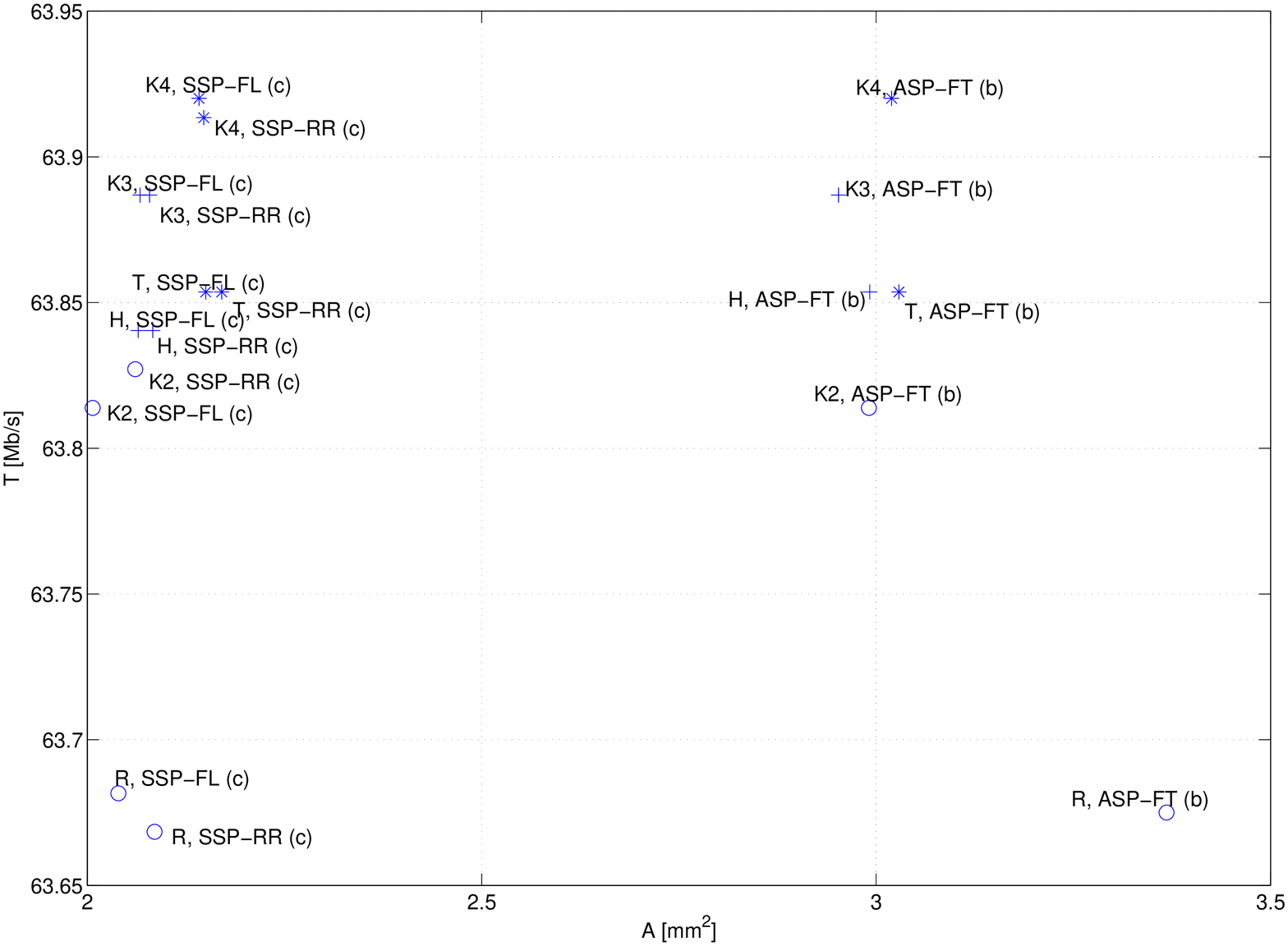}
    } \medskip
  \end{minipage}
\caption{Throughput/area comparison of different topologies and routing algorithm with DCM approach: 
WiMax interleaver, $N=2400$ for $R=1$, $P=64$ (a) and $R=0.33$, $P=16$ (b); circular shifting interleaver, 
$N=24576$ for $R=1$, $P=64$ (c) and $R=0.33$, $P=16$ (d)}
\label{fig:tar_tot}
\end{figure*}
\begin{table*}[t!]
  \centering
  \begin{threeparttable}[t]  
  \caption{Hardware resources breakdown for the circular shifting interleaver with $N$=24576, DCM approach: some significant points} 
  \label{tab:percentage}
  \begin{tabular}{|c|c|c|c|c|c|c|c|c|c|c|}
    \hline
$D$ & top. & $P$ & $R$ & routing alg. & Tot. FIFOs    & Tot. CB       & Tot. reg.     & RA/M          & IM+LM & Tot. \\
    &      &     &     &            & area [mm$^2$] & area [mm$^2$] & area [mm$^2$] & area [mm$^2$] & area [mm$^2$] & area [mm$^2$] \\
    \hline 
\rowcolor[gray]{0.9} 2   & R    & 64  & 1   & ASP-FT (b) & 11.45 (59.15\%) & 0.03 (0.15\%) & 0.08 (0.41\%) & 6.35 (32.80\%) & 1.45 (7.49\%) & 19.36 (100\%) \\
\rowcolor[gray]{0.8} 2   & R    & 64  & 1   & SSP-RR (c) & 27.46 (90.43\%) & 0.05 (0.16\%) & 0.14 (0.46\%) & 0.11 (0.36\%) & 2.61 (8.59\%) & 30.37 (100\%) \\
\rowcolor[gray]{0.7} 2   & R    & 8   & 0.33 & SSP-FL/RR (c) & 0.05 (2.84\%) & 0.01 (0.57\%) & 0.01 (0.57\%) & 0.01 (0.57\%) & 1.68 (95.45\%) & 1.76 (100\%) \\
\rowcolor[gray]{0.9} 2   & K    & 64  & 0.5 & ASP-FT (b) & 6.53 (60.46\%) & 0.03 (0.28\%) & 0.08 (0.74\%) & 2.71 (25.09\%) & 1.45 (13.43\%) & 10.80 (100\%) \\
\rowcolor[gray]{0.8} 2   & K    & 64  & 1 & SSP-RR (c) & 23.81 (89.11\%) & 0.05 (0.19\%) & 0.14 (0.52\%) & 0.11 (0.41\%) & 2.61 (9.77\%) & 26.72 (100\%) \\
\rowcolor[gray]{0.7} 2   & K    & 8  & 0.33 & SSP-FL (c) & 0.05 (2.86\%) & 0 (0\%)\tnote{(1)} & 0.01 (0.57\%) & 0.01 (0.57\%) & 1.68 (96.00\%) & 1.75 (100\%) \\
\hline
\rowcolor[gray]{0.9} 3   & H    & 64  &  1  & ASP-FT (b) & 9.25 (68.77\%) & 0.09 (0.67\%) & 0.11 (0.82\%) & 2.55 (18.96\%) & 1.45 (10.78\%) & 13.45 (100\%) \\
\rowcolor[gray]{0.8} 3   & H    & 64  &  1  & SSP-RR (c) & 18.02 (85.24\%) & 0.10 (0.47\%) & 0.18 (0.85\%) & 0.23 (1.09\%) & 2.61 (12.35\%) & 21.14 (100\%) \\
\rowcolor[gray]{0.7} 3   & H    & 8  &  0.33  & SSP-FL (c) & 0.05 (2.84\%) & 0.01 (0.57\%) & 0.01 (0.57\%) & 0.01 (0.57\%) & 1.68 (95.45\%) & 1.76 (100\%) \\
\rowcolor[gray]{0.9} 3   & K    & 64  &  1  & ASP-FT (b) & 7.86 (67.41\%) & 0.09 (0.77\%) & 0.11 (0.94\%) & 2.15 (18.44\%) & 1.45 (12.44\%) & 11.66 (100\%) \\
\rowcolor[gray]{0.8} 3   & K    & 64  &  1  & SSP-FL (c) & 13.15 (80.33\%) & 0.10 (0.61\%) & 0.18 (1.10\%) & 0.33 (2.02\%) & 2.61 (15.94\%) & 16.37 (100\%) \\
\rowcolor[gray]{0.7} 3   & K    & 8   & 0.33 & SSP-FL (c) & 0.06 (3.35\%) & 0.01 (0.56\%) & 0.02 (1.12\%) & 0.02 (1.12\%) & 1.68 (93.85\%) & 1.79 (100\%) \\
\hline
\rowcolor[gray]{0.9} 4   & T    & 64  & 1   & ASP-FT (b) &  5.14 (55.94\%) & 0.23 (2.5\%) & 0.13 (1.41\%) & 2.24 (24.37\%) & 1.45 (15.78\%) & 9.19 (100\%) \\
\rowcolor[gray]{0.8} 4   & T    & 64  & 1   & SSP-RR (c) &  13.21 (79.63\%) & 0.17 (1.02\%) & 0.23 (1.39\%) & 0.37 (2.23\%) & 2.61 (15.73\%) & 16.59 (100\%) \\
\rowcolor[gray]{0.7} 4   & T    & 8  & 0.33   & SSP-FL (c) & 0.06 (3.35\%) & 0.01 (0.56\%) & 0.02 (1.12\%) & 0.02 (1.12\%) & 1.67 (93.85\%) & 1.78 (100\%) \\
\rowcolor[gray]{0.9} 4   & K    & 64  & 1   & ASP-FT (b) & 3.78 (49.67\%) & 0.22 (2.89\%) & 0.13 (1.71\%) & 2.03 (26.68\%) & 1.45 (19.05\%) & 7.61 (100\%) \\
\rowcolor[gray]{0.8} 4   & K    & 64  & 1   & SSP-RR (c) & 6.86 (67.12\%) & 0.16 (1.57\%) & 0.23 (2.25\%) & 0.36 (3.52\%) & 2.61 (25.54\%) & 10.22 (100\%) \\
\rowcolor[gray]{0.7} 4   & K    & 8  & 0.33   & SSP-FL (c) & 0.06 (3.33\%) & 0.10 (0.56\%) & 0.02 (1.11\%) & 0.03 (1.67\%) & 1.68 (93.33\%) & 1.80 (100\%) \\
\hline
  \end{tabular}
  \begin{tablenotes}
  \item [(1)] The area and the percentage are not really zero, but they are negligible compared with the IM and LM 
contribution to the total area.
  \end{tablenotes}
  \end{threeparttable}
\end{table*}

The most important conclusions that can be derived from results in Table \ref{tab:wimax_results} and 
\ref{tab:mhoms_results} are:
\begin{enumerate}
\item The ASP-FT routing algorithm is the best performing solution both in terms of throughput and area 
when $R=1$.
\item The routing memory overhead of the ASP-FT algorithm (see Fig. \ref{fig:node} (b)) becomes 
relevant as $R$ decreases and SSP solutions become the best solutions mainly for $P=8$ and $P=16$.
\item In most cases topologies with $D$=4 achieve higher throughput with lower complexity overhead than topologies 
with $D$=2 when $R \to 1$.
\item In most cases, generalized de-Bruijn and generalized Kautz topologies are the best performing topologies.
\end{enumerate}
As a significant example, in Fig. \ref{fig:R1_ASP-FT}, 
we show the experimental results obtained with $R=1$ and ASP-FT routing algorithm 
for the WiMax interleaver with $N=2400$ (a) and the circular shifting interleaver with $N=24576$ (b).
Each point represents the throughtput and the area 
obtained for a certain topology with a certain parallelism degree $P$. Results referred to the same $P$ value are bounded 
into the same box and a label is assigned to each point to highlight the corresponding topology, namely 
topologies are identified as R-ring, H-honeycomb, T-toroidal mesh, K-generalized Kautz 
with the corresponding $D$ value (K2, K3, K4).

As it can be observed, 
generalized Kautz topologies with $D=4$ (K4) are always the best solutions to achieve high throughput 
with minimum area overhead.

In Fig. \ref{fig:tar_tot} significant results extracted from Table \ref{tab:wimax_results} and \ref{tab:mhoms_results}
are shown in graphical form. In particular, for $R=1$ the ASP-FT routing algorithm is the best solution, whereas for 
$R<1$ SSP routing algorithms, implemented as in Fig. \ref{fig:node} (c), tend to achieve 
the same performance as the ASP-FT routing algorithm with lower complexity overhead (see 
Fig. \ref{fig:tar_tot} (a) and (b) for the WiMax interleaver, $N=2400$ and Fig. \ref{fig:tar_tot} (c) and (d) for the circular shifting 
interleaver, $N=24576$).

An interesting phenomenon that arises increasing the interleaver size is the performance saturation that can 
be observed in the Table \ref{tab:mhoms_results} for $D=2$ topologies, namely 
the throughput tends to saturate 
and increasing $R$ has the effect of augmenting the area with a negligible increase or even with a decrease of throughput.
As an example, the generalized Kautz topology with $P=64$ and ASP-FT routing algorithm achieves more than 180 Mb/s 
with $R=1$, $R=0.5$, $R=0.33$. However, the solution with the smallest area is the one obtained with $R=0.33$.

The throughput flattening of low $D$ topologies can be explained by observing that high 
values of $R$ 
tend to saturate the network. Furthermore, high values of $R$ lengthen the input FIFOs as highlighted in Table 
\ref{tab:percentage}, where the total area of the network is given as the breakdown of the building blocks, 
namely the input FIFOs, the crossbars (CB), the output registers, the routing algorithm/memory (RA/M), the identifier 
memory (IM) and the location memory (LM) is given for some significant cases: 
the highest throughput (light-gray),
the highest area (mid-gray), 
and lowest area (dark-gray) points for each $D$ value in Table \ref{tab:mhoms_results}. 

\section{Conclusions}
\label{sec:concl}

In this work a general framework to design network on chip based turbo decoder architectures 
has been presented. The proposed framework can be adapted to explore different topologies,
degrees of parallelism, message injection rates and routing algorithms. Experimental results 
show that generalized de-Bruijn and generalized Kautz topologies achieve high throughput with a 
limited complexity overhead. Moreover, depending on the target throughput requirements different 
parallelism degrees, message injection rates and routing algorithms can be used to minimize the network 
area overhead. 
\bibliographystyle{IEEEtran}

\begin{thebibliography}{10}
\providecommand{\url}[1]{#1}
\csname url@samestyle\endcsname
\providecommand{\newblock}{\relax}
\providecommand{\bibinfo}[2]{#2}
\providecommand{\BIBentrySTDinterwordspacing}{\spaceskip=0pt\relax}
\providecommand{\BIBentryALTinterwordstretchfactor}{4}
\providecommand{\BIBentryALTinterwordspacing}{\spaceskip=\fontdimen2\font plus
\BIBentryALTinterwordstretchfactor\fontdimen3\font minus
  \fontdimen4\font\relax}
\providecommand{\BIBforeignlanguage}[2]{{%
\expandafter\ifx\csname l@#1\endcsname\relax
\typeout{** WARNING: IEEEtran.bst: No hyphenation pattern has been}%
\typeout{** loaded for the language `#1'. Using the pattern for}%
\typeout{** the default language instead.}%
\else
\language=\csname l@#1\endcsname
\fi
#2}}
\providecommand{\BIBdecl}{\relax}
\BIBdecl

\bibitem{wehn_TVLSI08}
T.~Vogt and N.~Wehn, ``Reconfigurable {ASIP} for convolutional and turbo
  decoding in an {SDR} environment,'' \emph{IEEE Transactions on VLSI},
  vol.~16, no.~10, pp. 1309--1320, Oct 2008.

\bibitem{berrou_ICC93}
C.~Berrou, A.~Glavieux, and P.~Thitimajshima, ``Near {S}hannon limit error
  correcting coding and decoding: {T}urbo codes,'' in \emph{IEEE International
  Conference on Communications}, 1993, pp. 1064--1070.

\bibitem{berrou_ITW01}
C.~Berrou, M.~Jezequel, C.~Douillard, and S.~Kerouedan, ``The advantages of
  non-binary turbo codes,'' in \emph{IEEE Information Theory Workshop}, 2001,
  pp. 61--63.

\bibitem{dobkin_TVLSI05}
R.~Dobkin, M.~Peleg, and R.~Ginosar, ``Parallel interleaver design and {VLSI}
  architecture for low-latency {MAP} turbo decoders,'' \emph{IEEE Transactions
  on VLSI}, vol.~13, no.~4, pp. 427--438, Apr 2005.

\bibitem{martina_TCASII08}
M.~Martina, M.~Nicola, and G.~Masera, ``A flexible {UMTS-WiMax} turbo decoder
  architecture,'' \emph{IEEE Transactions on Circuits and Systems II}, vol.~55,
  no.~4, pp. 369--373, Apr 2008.

\bibitem{giulietti_EL02}
A.~Giulietti, L.~V. der Perre, and M.~Strum, ``Parallel turbo coding
  interleavers: avoiding collisions in accesses to storage elements,''
  \emph{IET Electronics Letters}, vol.~38, no.~5, pp. 232--234, Feb 2002.

\bibitem{lee_EL02}
J.~Kwak and K.~Lee, ``Design of dividable interleaver for parallel decoding in
  turbo codes,'' \emph{IET Electronics Letters}, vol.~38, no.~22, pp.
  1362--1364, Oct 2002.

\bibitem{wehn_ISCAS02}
M.~J. Thul, N.~Wehn, and L.~P. Rao, ``Enabling high-speed turbo-decoding
  through concurrent interleaving,'' in \emph{IEEE International Symposium on
  Circuits and Systems}, 2002, pp. 897--900.

\bibitem{tarable_CL04}
A.~Tarable and S.~Benedetto, ``Mapping interleaving laws to parallel turbo
  decoder architectures,'' \emph{IEEE Communications Letters}, vol.~8, no.~3,
  pp. 162--164, Mar 2004.

\bibitem{polydoros_PIMRC08}
A.~Polydoros, ``Algorithmic aspects of radio flexibility,'' in \emph{IEEE
  International Symposium on Personal, Indoor and Mobile Communications}, 2008,
  pp. 1--5.

\bibitem{bougard_ICT08}
B.~Bougard, R.~Priewasser, L.~V. der Perre, and M.~Huemer,
  ``Algorithm-architecture co-design of a multi-standard {FEC} decoder
  {ASIP},'' in \emph{ICT Mobile Summit Conference}, 2008.

\bibitem{baghdadi_TVLSI09}
O.~Muller, A.~Baghdadi, and M.~Jezequel, ``From parallelism levels to a
  multi-{ASIP} architecture for turbo decoding,'' \emph{IEEE Transactions on
  VLSI}, vol.~17, no.~1, pp. 92--102, Jan 2009.

\bibitem{wehn_icecs02}
M.~J. Thul, F.~Gilbert, and N.~Wehn, ``Optimized concurrent interleaving
  architecture for high-throughput turbodecoding,'' in \emph{IEEE International
  Conference on Electronics, Circuits and Systems}, 2002, pp. 1099--1102.

\bibitem{wehn_icassp03}
------, ``Concurrent interleaving architectures for high-throughput channel
  coding,'' in \emph{IEEE International Conference on Acoustics, Speech and
  Signal Processing}, 2003, pp. 613--616.

\bibitem{speziali_EUROMICRO04}
F.~Speziali and J.~Zory, ``Scalable and area efficient concurrent interleaver
  for high throughput turbo-decoders,'' in \emph{IEEE Euromicro Symposium on
  Digital System Design}, 2004, pp. 334--341.

\bibitem{wehn_iscas05}
C.~Neeb, M.~J. Thul, and N.~Wehn, ``Network-on-chip-centric approach to
  interleaving in high throughput channel decoders,'' in \emph{IEEE
  International Symposium on Circuits and Systems}, 2005, pp. 1766--1769.

\bibitem{moussa_date07}
H.~Moussa, O.~Muller, A.~Baghdadi, and M.~.Jezequel, ``Butterfly and
  {B}enes-based on-chip communication networks for multiprocessor turbo
  decoding,'' in \emph{Design, Automation and Test in Europe Conference and
  Exhibition}, 2007, pp. 654--659.

\bibitem{moussa_iscas08}
H.~Moussa, A.~Baghdadi, and M.~.Jezequel, ``Binary de {B}ruijn interconnection
  network for a flexible {LDPC}/turbo decoder,'' in \emph{IEEE International
  Symposium on Circuits and Systems}, 2008, pp. 97--100.

\bibitem{baghdadi_EL06}
O.~Muller, A.~Baghdadi, and M.~Jezequel, ``Bandwidth reduction of extrinsic
  information exchange in turbo decoding,'' \emph{IET Electronics Letters},
  vol.~42, no.~19, pp. 1104--1105, Sep 2006.

\bibitem{bahl_TrIT94}
L.~Bahl, J.~Cocke, F.~Jelinek, and J.~Raviv, ``Optimal decoding of linear codes
  for minimizing symbol error rate,'' \emph{IEEE Transactions on Information
  Theory}, vol.~20, no.~3, pp. 284--287, Mar 1974.

\bibitem{benedetto_ETR98}
S.~Benedetto, D.~Divsalar, G.~Montorsi, and F.~Pollara, ``Soft-input
  soft-output modules for the construction and distributed iterative decoding
  of code networks,'' \emph{European Transactions on Telecommunications},
  vol.~9, no.~2, pp. 155--172, Mar/Apr 1998.

\bibitem{baghdadi_date06}
O.~Muller, A.~Baghdadi, and M.~Jezequel, ``{ASIP}-based multiprocessor {SoC}
  design for simple and double binary turbo decoding,'' in \emph{Design,
  Automation and Test in Europe Conference and Exhibition}, 2006, pp.
  1330--1335.

\bibitem{beniniNOC}
L.~Benini and G.~D. Micheli, ``Networks on chips: a new soc paradigm,''
  \emph{IEEE Computer}, vol.~35, no.~1, pp. 70--78, Jan 2002.

\bibitem{benini_date06}
L.~Benini, ``Application specific {NoC} design,'' in \emph{Design, Automation
  and Test in Europe Conference and Exhibition}, 2006, pp. 1330--1335.

\bibitem{vacca_DSD09}
F.~Vacca, H.~Moussa, A.~Baghdadi, and G.~Masera, ``Flexible architectures for
  {LDPC} decoders based on network on chip paradigm,'' in \emph{Euromicro
  Conference on Digital System Design, to appear}, 2009.

\bibitem{imase_TC81}
M.~Imase and M.~Itoh, ``Design to minimize diameter on building-block
  network,'' \emph{IEEE Transactions on Computers}, vol.~30, no.~6, pp.
  439--442, Jun 1981.

\bibitem{imase_TC83}
------, ``A design for directed graphs with minimum diameter,'' \emph{IEEE
  Transactions on Computers}, vol.~32, no.~8, pp. 782--784, Aug 1983.

\bibitem{parhami_TPDS01}
B.~Parhami and D.~M. Kwai, ``A unified formulation of honeycomb and diamond
  networks,'' \emph{IEEE Transactions on Parallel and Distributed Systems},
  vol.~12, no.~1, pp. 74--80, Jan 2001.

\bibitem{turbo_NOC_download}
M.~Martina, ``{T}urbo {NOC}: {N}etwork {O}n {C}hip based turbo decoder
  architectures,'' downloadable at www.vlsilab.polito.it/$\sim$martina.

\bibitem{SystemC}
``http://www.systemc.org.''

\bibitem{benedetto_EL96}
S.~Benedetto, D.~Divsalar, G.~Montorsi, and F.~Pollara, ``Algorithm for
  continuous decoding of turbo codes,'' \emph{IET Electronics Letters},
  vol.~32, no.~4, pp. 314--315, Feb 1996.

\bibitem{parhi_ISCAS04}
Y.~Zhang and K.~K. Parhi, ``Parallel turbo decoding,'' in \emph{IEEE
  International Symposium on Circuits and Systems}, 2004, pp. 509--512.

\bibitem{yao_ICASSP03}
A.~Abbasfar and K.~Yao, ``An efficient and practical architecture for high
  speed turbo decoders,'' in \emph{IEEE Vehicular Technology Conference}, 2003,
  pp. 337--341.

\bibitem{muller_ICCTA06}
O.~Muller, A.~Baghdadi, and M.~Jezequel, ``Exploring parallel processing levels
  for convolutional turbo decoding,'' in \emph{IEEE International Conference on
  Information and Communication Technologies: from Theory to Applications},
  2006, pp. 2353--2358.

\bibitem{dinoi_TCom05}
L.~Dinoi and S.~Benedetto, ``Variable-size interleaver design for parallel
  turbo decoder architectures,'' \emph{IEEE Transactions on Communications},
  vol.~53, no.~11, pp. 1833--1840, Nov 2005.

\bibitem{dolinar_TDA95}
S.~Dolinar and D.~Divsalar, ``Weight distributions for turbo codes using random
  and nonrandom permutations,'' \emph{TDA Progress Report}, vol. 42-122, pp.
  56--65, Aug 1995.

\end{thebibliography}

\end{document}